\documentclass{aa}     
\usepackage{graphics}

\begin{document}
   \thesaurus{07         
              (02-07.1;  
               02-18.8;  
               02-23.1;  
               08-19.4)  
}
\title{Statistics of the detection rates for
tensor and scalar gravitational waves from the
local galaxy universe}
\titlerunning{Rates of gravitational waves}
\author{Baryshev Yu.V. \inst{1}, Paturel G. \inst{2}}

\offprints{G. Paturel}

   \institute{
 Astronomical Institute of the Saint-Petersburg University,\\
 198504, St.-Petersburg, RUSSIA\\
\and
 CRAL-Observatoire de Lyon,\\
 F69561 Saint-Genis Laval CEDEX, FRANCE \\
}

   \date{Submitted 19 May 2000; Accepted 7 March 2001 }

   \maketitle

   \begin{abstract}
We use data on the local 3-dimensional galaxy distribution
for studying the statistics of the detection rates of gravitational waves (GW)
coming from supernova explosions.
We consider both tensor and scalar gravitational waves
which are possible in a wide range of relativistic and quantum
gravity theories.
We show that statistics of GW events as a function of sidereal time
can be used for distinction between scalar and tensor gravitational waves
because of the anisotropy of spatial galaxy distribution.

For calculation of the expected amplitudes of GW signals
we use the values of the released GW energy, frequency
and duration of GW pulse which are consistent with 
existing scenarios of SN core collapse.
The amplitudes of the signals
produced by Virgo and the Great Attractor
clusters of galaxies is expressed
as a function of the sidereal time
for resonant bar detectors operating now (IGEC) and for
forthcoming laser interferometric detectors (VIRGO).
Then, we calculate the expected number of  GW
events as a function of sidereal time
produced by all the galaxies within 100 Mpc.

In the case of axisymmetric rotational core collapse
which radiates a GW energy of $10^{-9}M_{\odot}c^2$, 
only the closest explosions can be detected.
However, in the case of nonaxisymmetric supernova explosion, due to 
such phenomena as centrifugal hangup, bar and lump formation,
the GW radiation could be as strong as that from a coalescing
neutron-star binary .
For radiated GW energy higher than $10^{-6}M_{\odot}c^2$ and  sensitivity
of detectors at the level $h \approx 10^{-23}$
it is possible to detect Virgo cluster and Great Attractor,
and hence to use the statistics of GW events for testing
gravity theories.

   \end{abstract}

\section{Introduction}

In a few years the third generation of gravitational
wave  detectors
will start searching for the most energetic
events in the Universe caused by gravitational collapse
and merging of relativistic  compact massive objects
(see the review by Thorne 1997).
This opens a new window onto the Universe and
creates new connections between optical extragalactic astronomy
and gravitational wave astronomy.
This will be the beginning of genuine observational
study of the physics of the
core collapse supernova explosions and testing
relativistic and even quantum gravity theories
(Damour 1999; Gasperini 1999).

Expected sources of powerful gravitational wave (hereafter, GW)
events are connected with
supernova explosions and merging of neutron stars and
other relativistic compact massive objects in galaxies.
Predicted GW signals essentially depend on the details
of the last relativistic stages of the gravitational collapse
which is still poorly known (Thorne 1987;1997; Paczyncki 1999; Burrows 2000).
Moreover studies of scalar-tensor gravity theories have shown that
spherical gravitational collapse and binary systems
generate scalar GWs which may be detected by existing GW detectors
(Baryshev 1982; 1995; 1997; Baryshev\& Sokolov 1984;
Sokolov 1992;
Damour\& Esposito-Fareze 1992; 1996; 1998;
Shibata et al. 1994; Harada et al. 1997;
Bianchi et al. 1998; Damour 1999; Brunetti et al. 1999;
Gasperini 1999; Novak\& Ibanez 1999; Maggiore\& Nicolis 2000
Nakao et al. 2000).

The aim of this paper is to estimate the contribution of nearby
galaxies and clusters of galaxies,
within a radius of about $100 h_{60}^{-1} Mpc$ around our Galaxy,
to the detection of the possible GW events.
This requires the knowledge of the actual 3-dimensional galaxy distribution
(section 2), the intrinsic rate of the most powerful events expected
from different galaxy types (section 3), and the amplitude
of the GW signal according to the prediction of  
existing  scenarios
of SN core collapses (section 4). 
In section 5 we calculate the probability of GW events
as a function of sidereal time for currently operating
bar detectors and forthcoming
interferometric detectors. In this section we first study the amplitude
expected for the Virgo cluster and the Great Attractor and then derive
the density of probability of GW events as a function of the sidereal
time for some detectors.
A discussion of the results and the main conclusions are given
in section 6.

\section{Distribution of galaxies within 100 Mpc}

Explosions of supernovae and mergings of binary massive compact objects
are very rare in our Galaxy. Hence, only observations of many galaxies
are expected to yield a reasonable detection rate.

\begin{figure*}
\resizebox{\hsize}{!}{\includegraphics{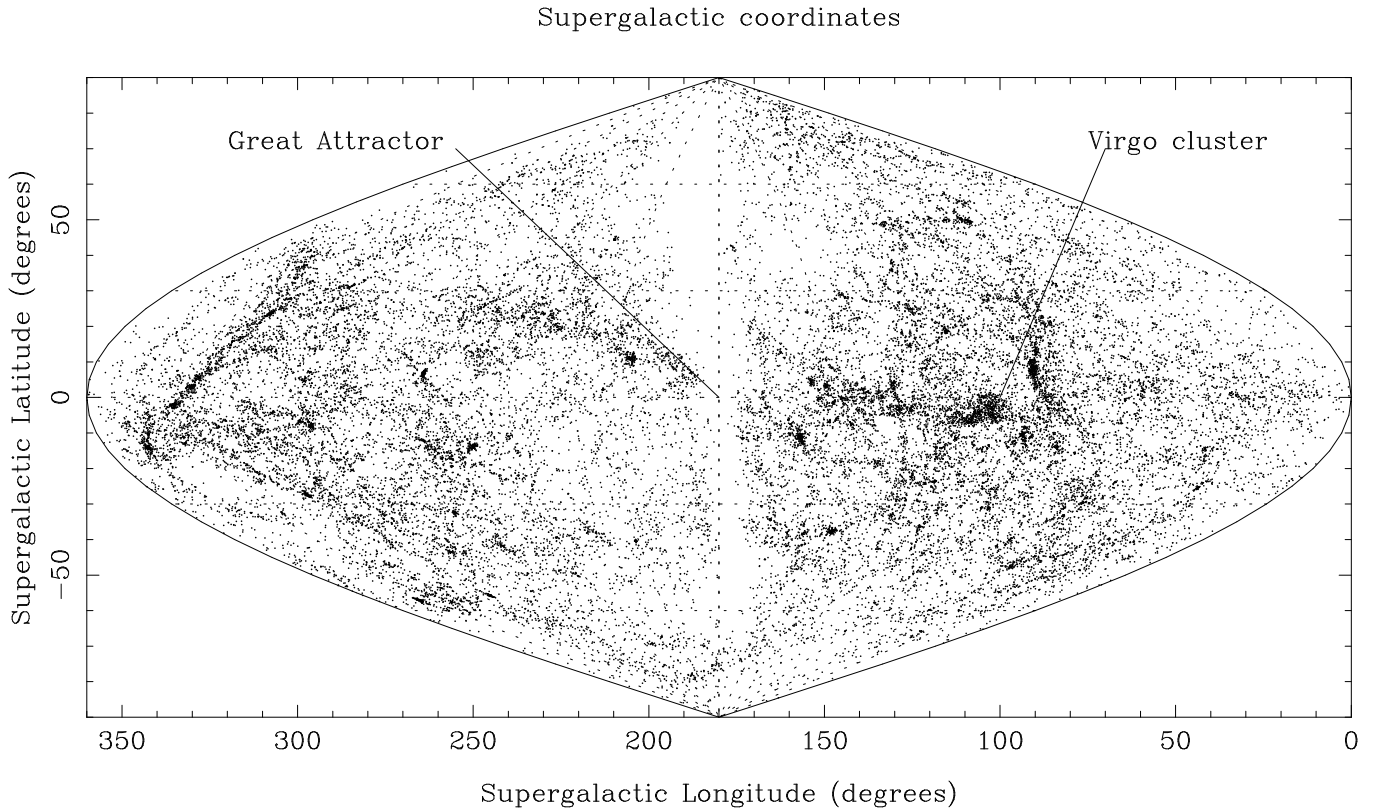}}
\caption{ Flamsteed equal area projection of our sample of 33557 galaxies
located within 100 Mpc. }
\label{flam}
\end{figure*}

From the Lyon-Meudon extragalactic database LEDA we extracted
a sample of 33557 nearby galaxies within 100 Mpc.
The 2D-distribution is shown on a Flamsteed equal area projection
(Fig. \ref{flam}). Some prominent structures appear. What are
their distances? What is the actual space density of galaxies?

\subsection{The problem of the determination of distances}
The direct determination of distance requires
difficult measurements and is thus available only for relatively
small samples (5000 galaxies). The most efficient method consists
of using the radial velocity together with a given Hubble constant.
However, the determination of the Hubble constant is still controversial.
Most of the difficulties come from the treatment of statistical biases.
From the most recent absolute calibration and the best unbiased 
determination we adopt:
\begin{equation}
H_o = 60 \pm 5 km.s^{-1}.Mpc^{-1}
\end{equation}

Nevertheless, we must consider that some results are still suggestive
of a larger value ($70 \pm 5 km.s^{-1}.Mpc^{-1}$),
and even $H_o = 50 km.s^{-1}.Mpc^{-1}$ cannot be excluded.
The smaller the value, the more difficult is GW detection
due to larger distances to GW sources.

\subsection{The problem of spatial distribution}

Analysis of the 3D-galaxy distribution from the correlation function method
(Davis \& Peebles,1983; Davis,1997) leads to the conclusion that the characteristic
correlation length is $l_0 \approx 5 Mpc$ and the maximum inhomogeneity scale
$l_{max} \approx 20 Mpc$. However, a more general statistical method to study
large-scale galaxy distribution has been recently developed
(see the review by Sylos Labini, Montuori, \& Pietronero, 1998). It is applicable to
any distribution of matter without the assumption of homogeneity
(which is required in the correlation function analysis).
The new analysis (the so-called conditional density function approach) is
actually taken from modern statistical physics where
it works as a standard tool.

Application of the conditional density function analysis to available
redshift surveys of galaxies, such as CfA, SSRS, Perseus-
Pisces, IRAS, LCRS etc, has revealed the fractal structure of the galaxy
distribution up to the scales corresponding to the depth of these catalogs,
i.e. about 100 Mpc (see e.g. Pietronero et al.,1997; Sylos Labini
et al.,1998). The fractal dimension of the spatial distribution
is close to $\mathcal{D}\approx 2$.
From the KLUN galaxy survey (Teerikorpi et al., 1998) where distances
to galaxies are obtained by the Tully-Fisher method, it was shown
that the observed number-distance relationship
corresponds to a fractal dimension $\mathcal{D} \approx 2.2$ and
continues up to the depth of the KLUN catalog, i.e. 200 Mpc.

The fractality implies that around any galaxy (including our own Galaxy)
the density decreases as $r^{\mathcal{D} - 3}$. This means that the
number of galaxies does not increase as $r^3$ but rather as $r^{\mathcal{D}}$.
The direct consequence for the present analysis is that the
detection rate of GW events will be lower than previously thought.

\begin{figure}
\resizebox{\hsize}{!}{\includegraphics{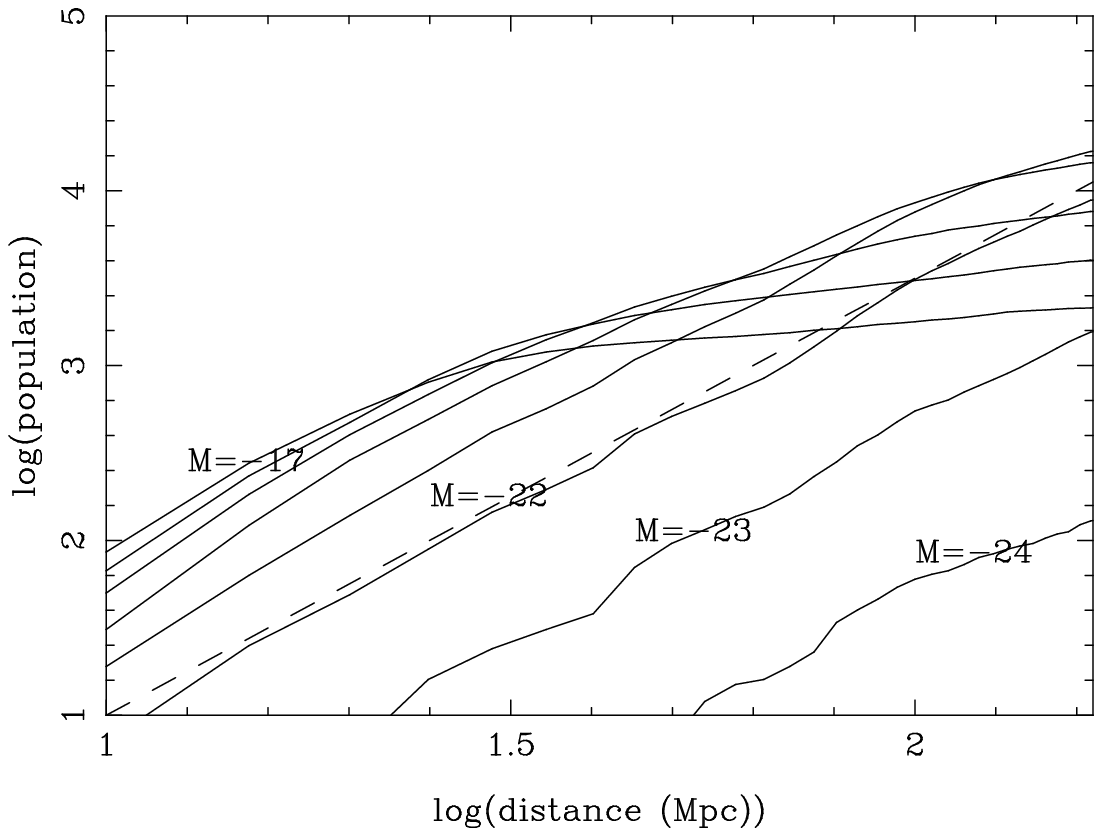}}
\caption{ Cumulative curves drawn for a wide range of absolute magnitudes
(from $M=-24$ to $M=-17$). The completeness is severely affected for
the less luminous galaxies ($M=-17$) when the distance increases. The
dashed curve shows the linear trend expected for a fractal dimension 2.5.
These curves are used to derive the true space density of galaxies.}
\label{complv}
\end{figure}

From our sample we plotted the cumulative curves $\log N$ vs. $\log r$ (N is the
total number of observed galaxies) within the radius $r$ (Mpc) (Fig. \ref{complv}).
 This is done for different absolute magnitudes. Intrinsically faint galaxies
($M=-17$) start to be missed beyond $ \log r=1.3$ ($\approx 20 Mpc$), while
galaxies brighter than $M=-22$ are observed up to the limit of our sample
($\approx 100 Mpc$).
The observed growth-curves correspond to $\mathcal{D} \approx 2.5$ (dashed curve in
Fig. \ref{complv}).
They are used to calculate the correction allowing us to estimate
the true number of galaxies in each direction, from the observed number.
At a given distance, this correcting factor is simply
deduced from the ratio of the observed and expected population (assuming
$\mathcal{D} \approx 2.5$).
It is to be noted that up to $\log r=1.3$, even the faintest galaxies ($M=-17$) follow
the linear curve. This means that the sample is complete up to this distance
(20 Mpc).

\section{Extragalactic sources of GW and intrinsic rates of events }

There are many sources of gravitational radiation in a galaxy. In fact,
any accelerated motion generates GW. Among those usually
discussed in the literature, galactic sources of GW are:
supernova explosions,
coalescing binary systems, binary stars, rotating asymmetric pulsars,
active galactic nuclei.
We consider only the GW sources which are expected to be
sufficiently frequent and efficient to be detected in the near future.

The most powerful sources of gravitational radiation are the core-collapse
supernovae (types Ib and II), merging neutron stars (ns), and black holes
 or other relativistic compact massive objects (cmo)
(Thorne 1987; 1997)) and also supernovae of type Ia, which are probably due to
the explosion of a CO white dwarf and might be sufficiently strong
candidates.

The relative supernova rates $R_{g,s}$
for galaxy type 'g' and SN type 's'
are free parameters in our code.
For further calculations we adopt these
from van den Bergh \& Tammann(1991), measured in SNU which equals
one event per 100 yr per $10^{10}L_0^B$.
In Table \ref{TRh}  we give the adopted rates of SN events
for different morphological types of galaxies.
Note that we use the Hubble constant 60 $km.s^{-1}Mpc^{-1}$ .

The event rate $R_{merg}$ for coalescing compact binaries composed
of ns (or cmo) is still widely discussed and has a large
uncertainty (e.g.Lipunov et al. 1997;
Portegies-Zwart\& McMillan 1999; Kalogera 1999).
Here we adopt the values from Lipunov et al.(1995), however these events
give a small contribution to the total statistics.

The first detailed study of the gravitational wave sky produced by
galaxies within 50 Mpc was done by Lipunov et al.(1995).
They considered a wide class of GW sources in galaxies and used
Tully's Nearby Galaxies Catalog comprising 2367 galaxies.
In this paper we use 33557 galaxies from the LEDA catalogue and
consider both tensor and scalar GW from supernova explosions.

\begin{table}
\caption{Adopted intrinsic rate of GW events. We consider Supernovae
explosions (SNIa, SNIb, SNII) and coalescences of White Dwarfs (WD),
Neutron Stars (NS), and Black Holes or Compact Massive Objects (CMO).
The rate may depend on the morphological type
of the parent galaxy (Elliptical(E), Spirals(S) or Irregulars(Irr)).
}
\label{TRh}
\begin{tabular}{lcrr}
\hline
 source  &  Rate $R_{g,s}$ of GW events        & \\
         &  per 100 years and $10^{10}L_{\sun}$ &             \\
\hline
SNIa     &  1.0 (E) 0.5 (S and Irr)  &        \\
SNIb     &  0.0 (E) 0.3 (S) 0.9 (Irr)&        \\
SNII     &  0.0 (E) 1.4 (S) 4.2 (Irr)&        \\
WD-WD    &       $3 \ 10^{-2}$       &        \\
NS-NS    &       $3 \ 10^{-3}$       &        \\
CMO-CMO  &       $    10^{-3}$       &        \\
\hline
\end{tabular}
\end{table}

\section{Theoretical amplitudes expected for GW events from SN }

\subsection{The problem of supernova explosion}

Expected amplitudes and forms of GW signals from supernova explosions
detected on the Earth by gravitational detectors
essentially depend on the adopted scenario of core-collapsed explosion
of massive stars and  relativistic gravity theory.
This is why the forthcoming GW astronomy
will give for the first time experimental limits
on possible theoretical models of gravitational collapse
including the strong field regime and even quantum nature
of the gravity force.

For the estimates of the energy, frequency and duration of
supernova GW emission
one needs a realistic theory of SN explosion
which can explain the observed ejection of the massive envelope.
Unfortunately, for the most interesting case of SNII explosion
such a theory does not exist now.
As was recently noted by Paczynski(1999), if there were no observations
of SNII it would be impossible to predict them from first
principles.

Modern theories of the core collapse supernova are
able to  explain all stages of evolution of a massive star
before and  after the explosion. However, the theory
of the explosion itself, which includes the relativistic stage of collapse
where a relativistic gravity theory
should be applied for the calculation of gravitational radiation,
is still controversial
and unable to explain the  mechanism by which the accretion shock
is revitalized into a supernova explosion (see the discussion by
Paczynski 1999; Burrows 2000).

Moreover, recent observations of
the polarization of core collapse supernovae
(Wang et al. 1999) and the relativistic jet in SN1987A
(Nisenson\& Papaliolios 1999) give strong evidence
in favor of a jet-induced explosion mechanism for massive supernovae
(Khokhlov et al. 1999; MacFadyen \& Woosley 1999; Wheeler et al. 2000).
Further evidence for highly asymmetric SN explosions comes from
recent observations of afterglows and host galaxies of gamma-ray
bursts (Paczynsky 1999).
This means that new "non-standard" scenarios of SN explosions
(and hence GW radiation)
may appear in the future and it may become important to study different
possibilities for the expected GW signal (hence a wide range
of GW parameters). 
Here we  adopt scenarios existing in the literature to
estimate the GW signal and we postpone the discussion of 
non-standard possibilities to section 6.

\subsection{Standard pulses of gravitational radiation}

The main aim of the present paper
was not to provide realistic supernova explosion models,
but to study the statistics of GW signals which may be expected
in standard SN explosion scenarios.
Hence we do not enter into the detailed calculations
of the precise forms of GW signals within
different gravity theories, but will simply use general energy arguments.

In our calculations we use the
standard pulse of gravitational radiation
introduced by Amaldi\& Pizzella(1979), which is a GW burst of
sinusoidal wave with amplitude $h_0$, frequency $\nu_0$
and duration $\tau_g$. For the case of tensor GW,
the amplitude $h_0$ of the signal
on the Earth due to the GW burst that occurs at a distance $r$
with total energy $E_{gw}$ is (see e.g. Pizzella 1989):

\begin{equation}
h_0 =1.4 \times 10^{-20} (\frac{1 Mpc}{r})(\frac{1 kHz}{\nu_0})
(\frac{E_{gw}}{1 M_{\odot}c^2})^{1/2}(\frac{1 s}{\tau_g})^{1/2}
\label{ho}
\end{equation}

For the case of scalar GW the amplitude $h_0$
is given by the same relation with a pre-factor
depending on the considered theory, e.g. for tensor field gravity theory
it is $\approx 2 \times 10^{-20}$ (Baryshev 1995).

Hence, each type of GW source at a fixed distance $r$ is
characterized by three main observable parameters
$E_{gw}$, $\nu_0$ and $\tau_g$.
In the next subsection we  choose acceptable values
for these parameters.

\subsection{ Adopted values for energy, frequency and duration of
tensor and scalar GW pulses}

Let us first consider the parameters for standard tensor GW pulses
in General Relativity. 
There is no unique widely accepted model
for tensor GW radiation produced by SN core collapse
and in the literature two main scenarios are usually discussed:  
axisymmetric and nonaxisymmetric ones, reviewed by Thorne (1997).

Within the theory of axisymmetric rotational core collapse, 
Zwerger \& Muller (1997) found that the energy spectrum
covers a frequency  $50 \ Hz < \nu_o < 3 \ kHz$
but most of the power is emitted
 between $500 \ Hz$ and $1 \ kHz$.
Duration of the pulses lies between $0.5 \ - \ 5 \ ms$. 
According to numerical calculations by Stark \& Piran (1985),
considered in Ferrari et al. (1999) as a basis for
prediction of GW background from SN, the GW energy spectrum
has two maxima around $5 \ kHz$ and $9 \ kHz$. 
The duration of GW pulses also is of the order a few $ms$.
In accordance with these calculations, for our statistical approach, 
we  adopt the characteristic frequency $\nu_o=1 \ kHz$
and the duration of the pulse 
 $\tau_g = 1 \ ms$.

For the total GW energy radiated by SN core collapse 
there is a very large range of predicted values in the literature.
According to  Zwerger \& Muller (1997) the energy
radiated in the form of GWs lies in the range
$6. \ 10^{-11}M_{\odot}c^2 < E_{gw} < 8. \ 10^{-8}M_{\odot}c^2$, 
which is consistent with Bonazzola \& Marck's (1993) results
for the deformation parameter $s < 0.1$.
However fully relativistic numerical simulations
by Stark \& Piran, also adopted by Ferrari et al. (1999),
give $E_{gw} \leq  10^{-3}M_{\odot}c^2$.
Moreover, if the collapsing core rotates so rapidly that
it becomes nonaxisymmetric and may be transformed into
a bar-like configuration, which also  might break up into  
several fragments, then the GW radiation could be almost
as strong as that from a coalescing neutron star binary.
Several specific scenarios for such nonaxisymmetric SN
core collapses have been proposed (see review by Thorne,1997).
According to Lai \& Shapiro (1995) the energy radiated
in GW during the nonaxisymmetric stage of the gravitational
collapse can be as large as $4\times 10^{-3}M_{\odot}c^2$.
Bonnell \& Pringle (1995) considered gravitational radiation
from SN core collapse and fragmentation, which could produce
the GW energy $10^{-2}M_{\odot}c^2$. 
For our statistical study of GW events we choose, as a basis,
the value $E_{gw} = 10^{-6}M_{\odot}c^2$.

Within classical general relativity there is no scalar GW
and for instance spherically symmetric collapse does not
generate gravitational radiation.
However other relativistic and quantum gravity theories
predict both tensor and scalar GW. Calculations
of amplitudes, frequencies and forms of the scalar
gravitational radiation in the case of spherically
symmetric SN core collapse has been made by
Shibata et al.(1994), Harada et al.(1997), Novak \& Ibanez (1999).
The released scalar GW energy is of order $\omega ^{-1}M_{\odot}c^2$,
where $\omega \geq 10^3 $ is the parameter of the Brans-Dicke theory.
The characteristic frequency and duration are similar
to the tensor GW in general relativity.
For the comparison of the statistics of tensor and scalar events
we adopt the same energy, frequency and duration for both scalar
and tensor GW's.

\begin{figure}
\resizebox{\hsize}{!}{\includegraphics{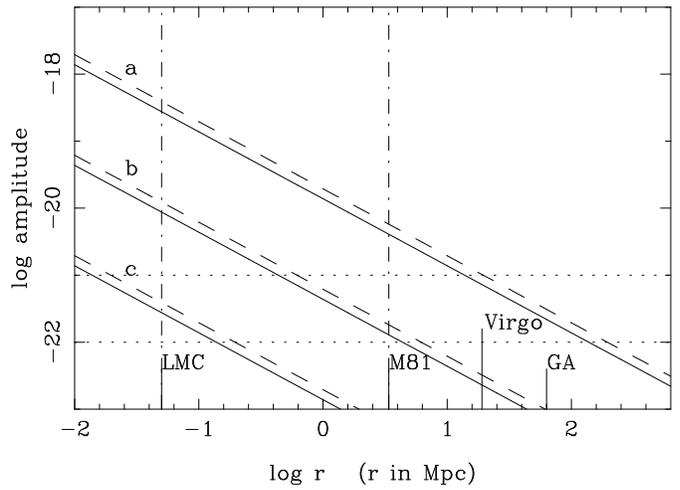}}
\caption{Theoretical GW amplitude versus distance. The predicted amplitude at
a distance $r$ is given for tensor (solid lines) and scalar (dashed curves)
waves according to Eq.2 for cases a,b,c (see text).
For instance, case $b$ corresponds to
the GW energy of $10^{-6}$ $M_{\odot }c^2$, a pulse duration
$\tau_g =10^{-3} \ s$ and a frequency $\nu_0 =1 \ kHz$.
Two sensitivity levels are represented (dotted lines) corresponding
to $h_0 = 10^{-21}$ and $10^{-22}$. The distances
of LMC, M81, Virgo and Great Attractor are labelled on the x-axis.}
\label{sens_dist}
\end{figure}

In Fig.\ref{sens_dist} we plot the expected amplitudes of GWs
calculated according to Eq.2 for a wide range of the 
parameter $E_{gw}$, which covered the expected values of GW energy 
for tensor and scalar GW. Three pairs of lines (a,b,c)
correspond to the following combinations of the main GW 
pulse parameters:\\
a)$E_{gw}=10^{-3} M_{\odot}c^2$; $\nu_0 = 10^3$Hz; $\tau_g =1 ms$;\\
b)$E_{gw}=10^{-6} M_{\odot}c^2$; $\nu_0 = 10^3$Hz; $\tau_g =1 ms$;\\
c)$E_{gw}=10^{-9} M_{\odot}c^2$; $\nu_0 = 10^3$Hz; $\tau_g =1 ms$.\\

For further calculations we adopt the values of GW amplitudes which
correspond to the case $b$.
Obviously, our calculations may be rescaled by using any
other combinations of main GW parameters according to Eq.2.

In Figure \ref{sens_dist} we draw two levels of sensitivity 
(horizontal dotted lines) which can be expected today 
($h_0 = 10^{-21}$ and $10^{-22}$).
One sees that it would be possible to detect SN explosions at the 
distance of the Virgo cluster and of the Great Attractor only
with an optimistic energy release (case $a$).

\subsection{Longitudinal and transversal scalar GW}

An important difference between tensor and scalar GWs is that tensor
waves (spin 2) are transversal while scalar waves (spin 0)
can be transversal and/or longitudinal.
There is no scalar GW in general relativity.

In the frame of the Jordan-Fierz-Brans-Dicke theory,
as for any metric tensor-scalar theories,
the scalar wave is transversal but isotropic in the plane
transversal to the propagation direction
(Damour \& Esposito-Farese (1992).
The transversal spin 0 wave may be presented as the metric
perturbation in the following form
(Bianchi et al. 1998;  Brunetti et al. 1999; Maggiore \& Nicolis 2000):
\begin{equation}
g_{ik} = h_0 \cos(\omega t - kz) diag(0,1,1,0)
\label{trans}
\end{equation}

In the frame of the field approach to gravity 
formulated in 60's by Feynman
(see e.g. Feynman et al. 1995; Thirring 1961;  Straumann 2000), 
there are also scalar gravitational waves which are
longitudinal and generated by the trace
of the energy-momentum tensor (Baryshev 1982,1995; Sokolov 1992). 
In such a scalar wave, a test particle moves
along the direction of the wave propagation.
The most straightforward way to demonstrate this is to use
exact relativistic equations of motion of test
particles in tensor and scalar potentials derived by Kalman (1961).
It is an important fact that the physical interaction
of a GW with a detector may be completely analyzed in terms
of the weak field approximation, i.e. with the usual Lagrangian formalism
of relativistic field theory in Minkowski space
(see Appendix).

The scalar plane monochromatic GW in the system of coordinates
with the z-axis directed along the wave propagation may be presented in
the form:
\begin{equation}
h_{ik} = h(t,z)\eta_{ik} 
\end{equation}
where $h(t,z)$ is a 4-scalar field of Minkowski space, or
\begin{equation}
h_{ik} = h_0 \cos(\omega t - kz) diag(1,-1,-1,-1)
\label{confo}
\end{equation}
where $h_0\ll c^2$ is the amplitude of the wave,
$\omega = 2\pi \nu_0$, $k =\omega/c = 2\pi/\lambda$,
$\eta_{ik} = diag(1,-1,-1,-1)$ is the Minkowski metric tensor.
As shown in the Appendix the test particles in such a gravitational
potential are moving in the direction of the wave propagation
(z-axis). The analysis of the motion of test particles
is sufficient for the description of an interaction
of the scalar GW with interferometric and bar detectors because
the scalar GW  does not interact with electromagnetic field.
Indeed, the interaction Lagrangian is $\Lambda_{int}=h_{ik}T_{em}^{ik}=0$
due to the tracelessness of the energy-momentum tensor of the
electromagnetic field. Hence without any ambiguity, the scalar GW
is longitudinal in the field gravity theory.

This means that both
types of scalar GWs, longitudinal and transversal, are theoretically
possible and physically different.
The physical difference between  longitudinal and  transversal
scalar GW may be experimentally established by comparing the direction
of the maximum sensitivity of a bar detector with the direction
of the axis of the bar.
Indeed, for a bar detector
the maximum sensitivity to the transversal GW is in the direction
orthogonal to the bar axis, while for the longitudinal scalar GW the
maximum sensitivity direction is along the axis of the bar.

For an interferometric detector (which has two arms) the maximum
sensitivity to the tensor GW will be in the direction orthogonal
to the plane containing both arms. For the longitudinal
scalar waves there are two directions of maximum sensitivity which coincide
with the directions of each arm. It is interesting that for the
transversal scalar GW the directions of maximum sensitivity also
coincide with the two arms of the interferometer
(Maggiore,Nicolis 2000; Nakao et al.2000).
This is a special
case of a two arm interferometer, while in the case of a bar detector
the detector patterns are different for longitudinal and transversal
scalar GWs.

The difference in the response of GW detectors to arriving
tensor and scalar GW pulses
allows one to test the nature of the detected waves.
Below we compare the statistics of expected GW events
for transversal and longitudinal GWs.

\section{Probability of GW events as a function of sidereal time}

\subsection{The expression for the number of GW events}
Let us consider a sample of galaxies. Each galaxy may produce GW events with a certain
rate. These events will be either detected or not, depending on the amplitude of the
signal and on the sensitivity of the detector.
The number of detected events produced at a time $t$ for a given type of wave
(tensor or scalar noted by index $w$) and a given
detector (interferometer or bar detector noted by index $d$) will be the
sum of the contribution
of each galaxy to the considered type of source (SNIa, SNIb, SNII etc... noted by index $s$).
\begin{equation}
\frac{dP}{dt}(t, s, w, d)= \sum_{g} L_g \times  R_{g,s} \times  \Delta
\label{proba}
\end{equation}
where $L_g$ is the luminosity  (in $L_{\odot}$) of the g-th galaxy;
$R$ is the rate of GW events per year and per $L_{\odot}$
for the considered type of source and the considered type of wave ; $\Delta$ is the detectability  calculated from
the observed amplitude $h_{obs}$ and the limiting amplitude $h_{lim}$
of the detector.

Let us detail the three terms used in Eq. \ref{proba}.

\begin{itemize}
\item[$\bullet$] The luminosity $L_g$ is derived from the absolute blue magnitude $M_{g}$ of the g-th
galaxy. A correction for incompleteness is done using  Fig.\ref{complv} and assuming that
missed galaxies are distributed like the observed ones. This does not
account for structures completely hidden by the disk of our Milky Way 
\footnote{For example,
the Great Attractor which is presented in section 5.2}. The luminosity is:
\begin{equation}
L_{g}= C(M_{g},r_g) 10^{-0.4(M_{g}-M_{\odot})}
\end{equation}
($r_g$ is the distance of the g-th galaxy and $M_{\odot}=5.2$ is the absolute magnitude of
the Sun in the photometric B-band)
From Fig.\ref{complv} it is visible that the correcting term $C$ is almost equal to one
below 20Mpc ($\log r =1.3$), because all curves are linear even for the faintest
galaxies. This means that up to the Virgo cluster ($\approx 18$Mpc) all galaxies are
included.
\item[$\bullet$] The rate $R_{g,s}$ depends on the morphological type of the considered galaxy (index $g$)
and on the type of the GW source (index $s$) according to Table \ref{TRh}.
As an example, we made calculations for SNII + SNIb.
Hence, the expected counts will be
underestimated, if all other conditions are satisfied.
\item[$\bullet$] The detectability $\Delta$ is defined as:
\begin{equation}
 \Delta= \left\{ \begin{array}{ll}
 1  & \mbox{ if $h_{obs} > h_{lim}$},\\
 0  & \mbox{ if $h_{obs} < h_{lim}$}.
\end{array}
\right.
\end{equation}
We will consider a sensitivity of $h_{lim}=10^{-23}$. It is
not yet accessible by present and scheduled detectors, but one can 
hope an improvement will be possible from space missions.
The observed amplitude is defined as:
\begin{equation}
h_{obs}= h_o .  | G(\alpha, \delta, w,d) |
\label{eq13}
\end{equation}
$h_o$ is given by Eq.\ref{ho}. It depends on the considered type of source (via the energy, frequency and
duration of the pulse) and on the considered type of gravitational wave. $ G(\alpha, \delta, w,d)$
is the geometrical factor. It characterizes the relative orientation of the GW and of the detector at the time
of the observation. It depends on the type and position of the detector (characterized
by its latitude, sidereal time and azimuth of its reference axis), on the direction of the GW
(characterized by the equatorial coordinates
$\alpha, \delta$ of the g-th galaxy), and on the type of wave
(including polarization for tensor waves).
Note that the longitude of the site is not needed explicitly because it is included in the
definition of the sidereal time. The sidereal time will be given in
hours (from $0$ to $24$).
\end{itemize}

The calculation of the geometrical factor $G$ is explained in the next subsection for
interferometric and bar detectors.

\subsection{Geometry of the system}
\begin{figure}
\resizebox{\hsize}{!}{\includegraphics{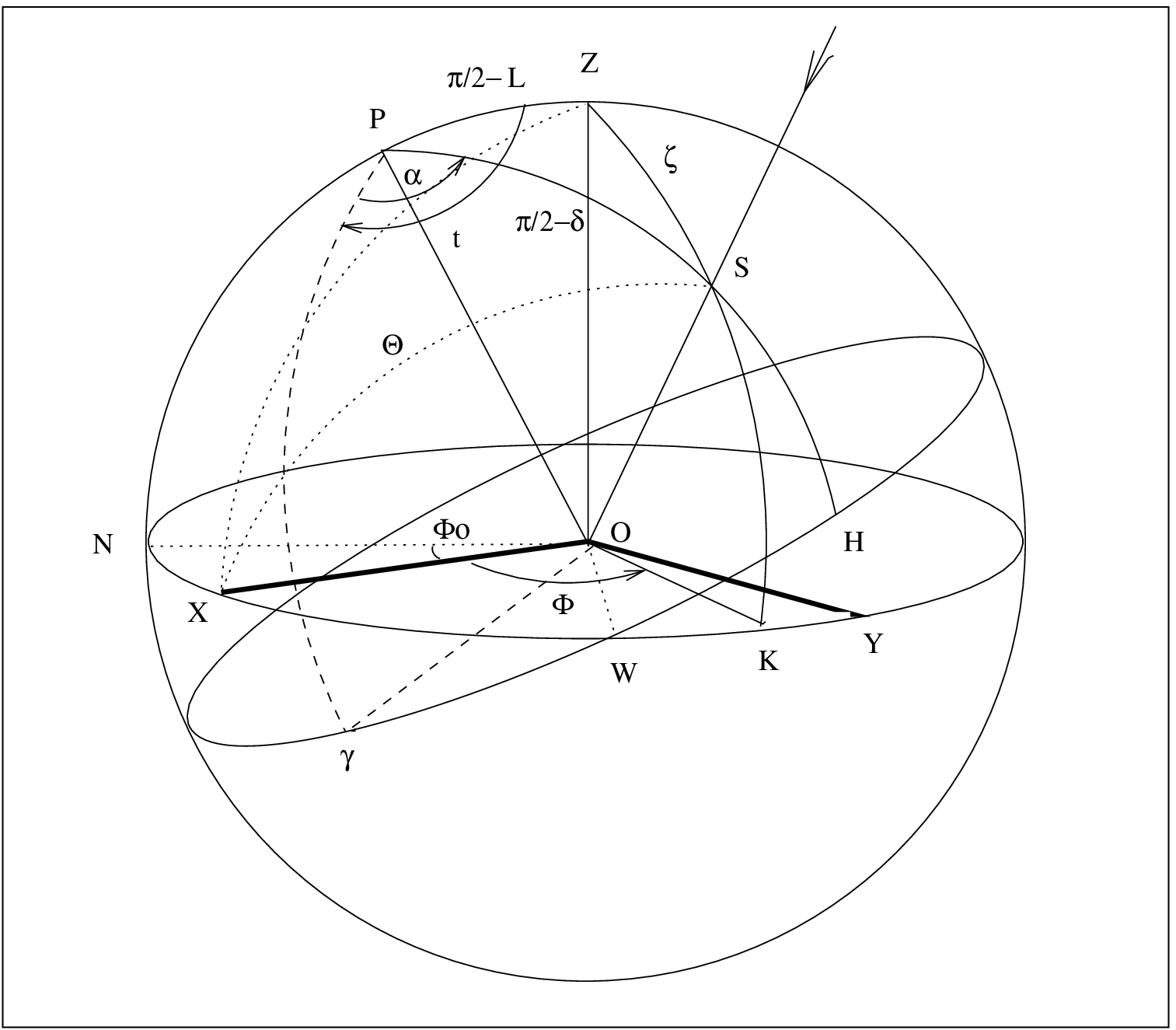}}
\caption{The main geometrical definitions. $Z$ is the zenith of the site. $P$
is the northern pole of the equatorial coordinate system. $\gamma$ defines the
sidereal time with respect to the southern meridian.
The source $S$ is located by its equatorial coordinates  $\alpha=\widehat{\gamma O H}$
for the right ascension and $\delta = \widehat{HOS}$ for the declination. The
distance to the zenith is $\zeta=\widehat{ZOS}$. The reference
direction for the detector is the direction $OX$. It is either the x-arm for an interferometric
detector or the direction of the bar for a bar detector. The azimuth of this direction is
$\Phi_o$. It is defined in the direct sense from the north to the west.}
\label{GWgeo}
\end{figure}

The geometrical configuration is defined by the wave direction ($OS$),
the local horizon (perpendicular to $OZ$, where $Z$ is the zenith) and
the equatorial plane (perpendicular to $OP$, where $P$ is the equatorial
northern pole). The main features of the system are shown in Fig.\ref{GWgeo}.

The expression of the geometric factor for tensor waves has been calculated
by Schutz \& Tinto (1987) and Thorn (1987). For scalar waves it has been calculated
by Baryshev (1997).

The relevant angles are $\zeta$, $\Phi$, $\Theta$ and $\Psi$.
They depend on the different cases. Let us detail  :
\begin{itemize}
\item[$\bullet$] For interferometric detectors: the zenith
distance $\zeta=\widehat{ZOS}$ (angle between the direction of the zenith and the
arrival-direction of the wave),
the azimuth $\Phi=\widehat{XOK}$ of the direction of the wave (direction OK) calculated with respect to
the reference axis of the detector.
\item[$\bullet$] For bar detectors: only the angle $\Theta=\widehat{XOS}$ is relevant.
\item[$\bullet$] For tensor GW the polarization angle $\Psi$. As in Schutz \& Tinto (1987)
it may be
measured in the GW plane from the line of nodes (intersection of the GW plane and of
the horizontal plane).
\end{itemize}

Note that the direction of the polarization $\Psi$ depends on the geometry
of the emitting source. For the calculation with the actual
galaxy distribution $\Psi$  will be simply chosen at random.
With these notations we obtain the following results:
$\zeta$ and $\Phi$ are calculated by solving the spherical triangle $ZPS$.
One obtains:
\begin{equation}
\cos \zeta = \sin L \sin \delta + \cos L \cos \delta \cos (t- \alpha)
\end{equation}
$L$ is the latitude of the site where the detector operates.
 $\alpha$ and $\delta$ are the equatorial coordinates of the source, and $t$ is
the sidereal time at the site. $\zeta$ is defined in the range ($0,\pi$).

\begin{equation}
\cos (\Phi_o + \Phi) = \frac {\cos L \sin \delta - \sin L \cos \delta \cos (t- \alpha)}{\sqrt{1- \cos^2 \zeta}}
\end{equation}
\begin{equation}
\sin (\Phi_o + \Phi) = \frac {\cos \delta \sin (t- \alpha)}{\sqrt{1- \cos^2 \zeta}}
\end{equation}
$\Phi_o$ is the azimuth of the reference axis of the system counted in the direct
sense from the direction of the north. In this paper the reference axis is $OX$, 
as in Thorn 1987. It is either the direction of the X-arm for 
an interferometric detector or the direction of the bar for a bar detector.
$\Phi$ is defined over the range ($0,2 \pi$).

The angle $\Theta$ is calculated by solving the triangle $XZS$:
\begin{equation}
\cos \Theta = \sin \zeta \cos \Phi
\end{equation}
$\Theta$ is defined over the range ($0, \pi$).

The relevant angles being calculated according to the previous relations, the geometrical factor
in the four considered cases has the following form:

{\bf 1}. For tensor waves and interferometric detector the geometrical factor is:

\begin{equation}
\label{relti}
G = 0.5(1+\cos^2 \zeta) \cos 2 \Phi  \sin 2 \Psi - \cos \zeta \sin 2 \Phi \cos 2 \Psi
\end{equation}

{\bf 2}. For longitudinal scalar waves and interferometric
detector the geometrical factor is:

\begin{equation}
\label{relsi}
G = \sin \zeta (| \cos \Phi | - | \sin \Phi |)
\end{equation}

{\bf 3}. For tensor waves and bar detector the geometrical factor is:

\begin{equation}
\label{reltb}
G = \sin^2 \Theta \cos 2 \Psi
\end{equation}

{\bf 4}. For longitudinal scalar waves and bar
detector the geometrical factor is (Appendix, Eq.\ref{dl/l}):

\begin{equation}
\label{relsb}
G = \cos \Theta
\end{equation}

The interferometric GW detectors (such as TAMA, GEO600, VIRGO, LIGO)
have the frequency range of about $10 - 1000 Hz$ and the sensitivity
$h \approx 10^{-20} - 10^{-22}$. Sensitivity  is a measure
of the detectable amplitude and it is proportional to $\Delta l /l$,
the relative length variation of the arms of the interferometer.

Presently working bar detectors, such as IGEC RBO (International
Gravitational Event Collaboration of Resonant Bar Observatory)
which includes five cryogenic resonant bar detectors
(ALLEGRO, AURIGA, EXPLORER, NAUTILUS, NIOBE), have a typical
bandwidth of the order of 1 Hz around each one of the two
resonances (close to 1000 Hz). The achieved sensitivity is now
$F_0 \approx 2-6 \times 10^{-21} Hz^{-1}$ (Prodi et al. 2000).
In future one expects the sensitivity to be at the level of $10^{-22} Hz^{-1}$.

We have calculated the predicted amplitude for different existing
detectors and specific galaxy clusters.
The closest concentration of galaxies is the Virgo cluster. The name
of the VIRGO detector comes from the name of the cluster itself because
it may be the main source of first detectable GW events. The Virgo cluster
is not very far from our Local Group (20 Mpc). It induces a velocity
of about 170 $km.s^{-1}$ on the Local Group. On the other hand, it has
been claimed (Dressler et al., 1987) that a hidden large concentration of
galaxies (hereafter, the Great Attractor) induces a velocity
of about 500$km.s^{-1}$ on our Local Group (i.e. about three times more).
The distance of such a concentration has been estimated to be three times
the distance to Virgo. This means that the number of galaxies could be about
27 times larger than the number of Virgo galaxies
 \footnote{The velocity, and thus
the acceleration during the same time, is three times larger. The distance
is also three times larger, resulting in a mass nearly $3 / 3^{-2}=27$
times larger}.
If the sensitivity of GW detectors is improved, the Great Attractor may become
the major source of GW events. This justifies the interest we place in this
region. The position of this putative Great Attractor would be
roughly at galactic coordinates $l=310 \deg$,
$b=0 \deg$. This is well supported by the apparent 2D-distribution
of galaxies which shows that this region may constitute a link
between two visible structures on both sides of the Milky way
(Paturel et al., 1987 and Fig.\ref{flam}) and by the
discovery of a large number of galaxies around this region
(Kraan-Korteweg, 2000).

Then, we considered two clusters as dominant sources.
The adopted equatorial coordinates and distances are the following:\\

\vspace{0.3cm}
\begin{tabular}{llll}
Cluster         & $\alpha$(1950) &  $delta$(1950)  & $r$ \\
\hline
Virgo           &$12h 28 m$&$+12 \deg 40 '$ &$20 Mpc$\\
Great-Attractor &$15h 00 m$&$-60 \deg 00 '$ &$60 Mpc$\\
\hline
\end{tabular}
\vspace{0.3cm}

We considered the following detectors:\\

\vspace{0.5cm}
\begin{tabular}{llrl}
Detector        & Latitude $L$&  Azimuth $\Phi_o$  & Type \\
\hline
VIRGO           &$44.7\deg $&$ -15\deg $& interf.\\
AURIGA          &$45.4\deg $&$ +45\deg $& bar    \\
NAUTILUS        &$41.8\deg $&$   0\deg $& bar    \\
NAUTILUS        &$41.8\deg $&$  90\deg $& bar    \\
\hline
\end{tabular}
\vspace{0.5cm}

In the following subsections we calculate the amplitudes in different
conditions, for different detectors and sources.
For the study of the polarization effect on tensor waves
(Fig.\ref{ampliTIi} to \ref{ampliTIa}) we used the cluster Virgo (as
a point source) and detectors VIRGO and AURIGA. For the comparison
of the distribution of the amplitudes of Virgo and of the Great-Attractor
(Fig. \ref{ampliTIm} to \ref{ampliSB2m}) we used Virgo and the Great
attractor as point sources and the detectors VIRGO, AURIGA and NAUTILUS.
For the calculation of the density of probability of GW events 
along the sideral time (Fig. \ref{COUNT_A} and \ref{COUNT_C}) we
used the sample of individual galaxies and the detectors VIRGO, 
AURIGA and NAUTILUS.

\subsection{Effect of the polarization}
\begin{figure}[!]
\resizebox{\hsize}{!}{\includegraphics{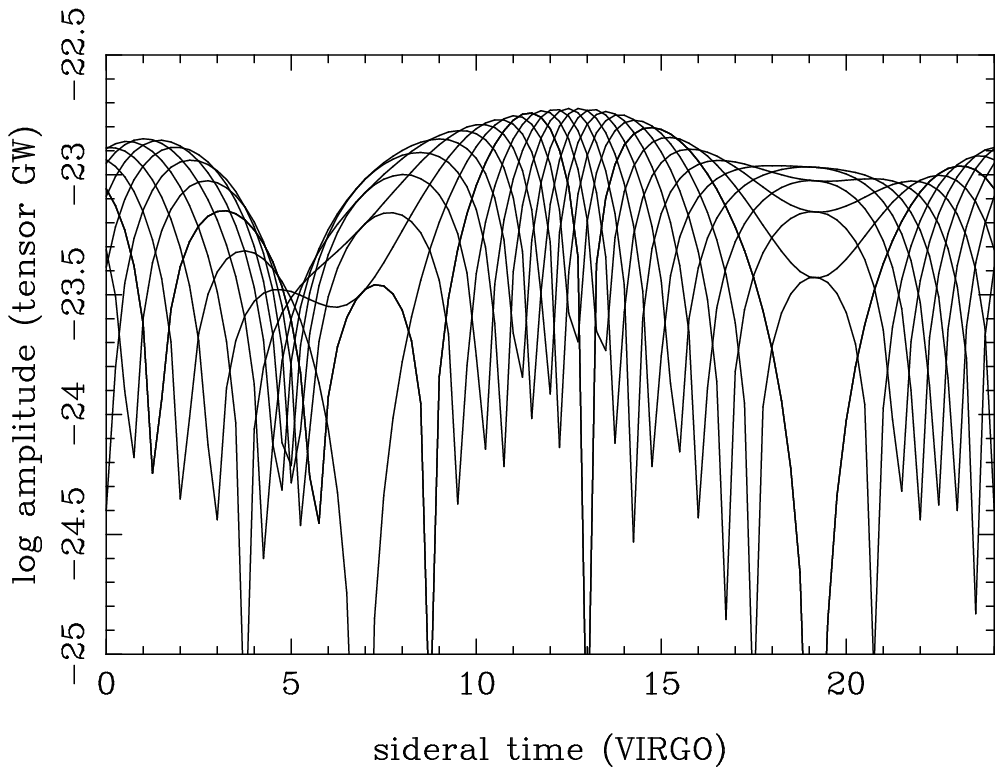}}
\resizebox{\hsize}{!}{\includegraphics{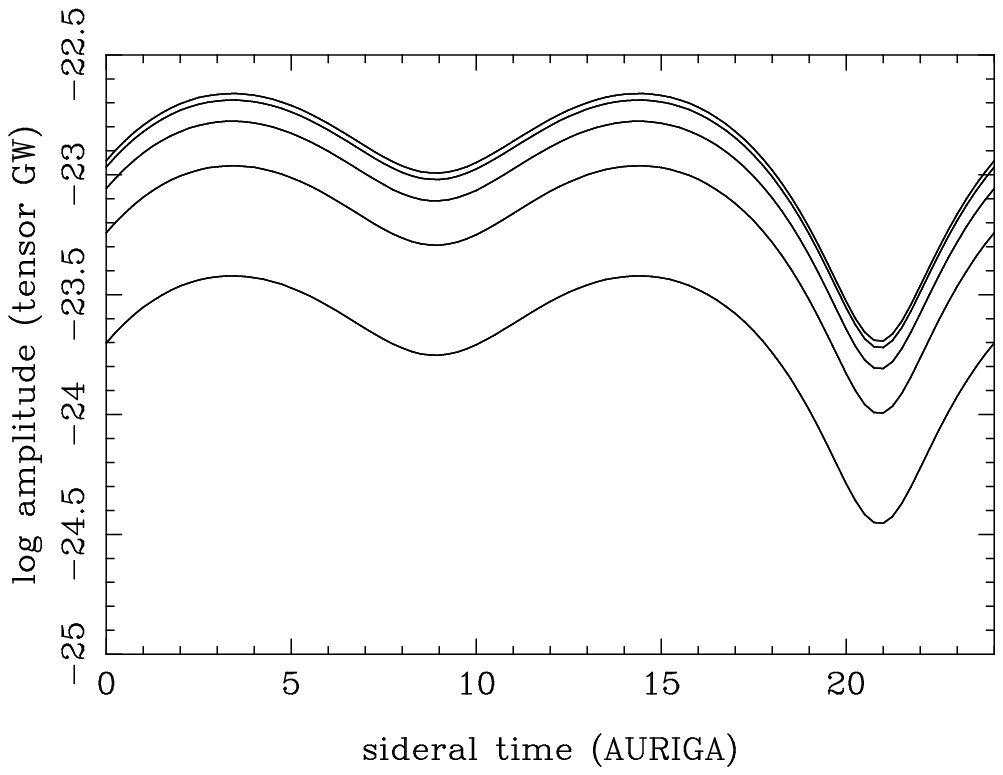}}
\caption{a) Amplitude as a function of sidereal time for tensor GW emitted
by Virgo and seen with
the VIRGO interferometric detector for different polarizations of the GW.
Each curve represents 36 fixed polarization directions.
b) Amplitude as a function of sidereal time for tensor GW emitted
by Virgo and seen with
the AURIGA bar detector for different polarizations.
}
\label{ampliTIi}
\end{figure}

Because the polarization angle cannot be predicted,
it is important to show the effect of the polarization for tensor waves.
Considering only the Virgo cluster and interferometric (VIRGO)
and bar detector (AURIGA),
we calculated the amplitude according to Eq. \ref{eq13} using $h_o$
for case b (Fig.\ref{sens_dist}). The geometrical factor is given
by Eq. \ref{relti} and \ref{reltb}, for an interferometric and bar detector
respectively. We used
36 polarization angles over the range $(0,2\pi)$ \footnote{ In fact,
the period is $\pi /2$ for interferometric detectors and $\pi /4$ for bar detectors}.
The results are shown in Fig. \ref{ampliTIi}a-b , respectively.

\begin{figure}[!]
\resizebox{\hsize}{!}{\includegraphics{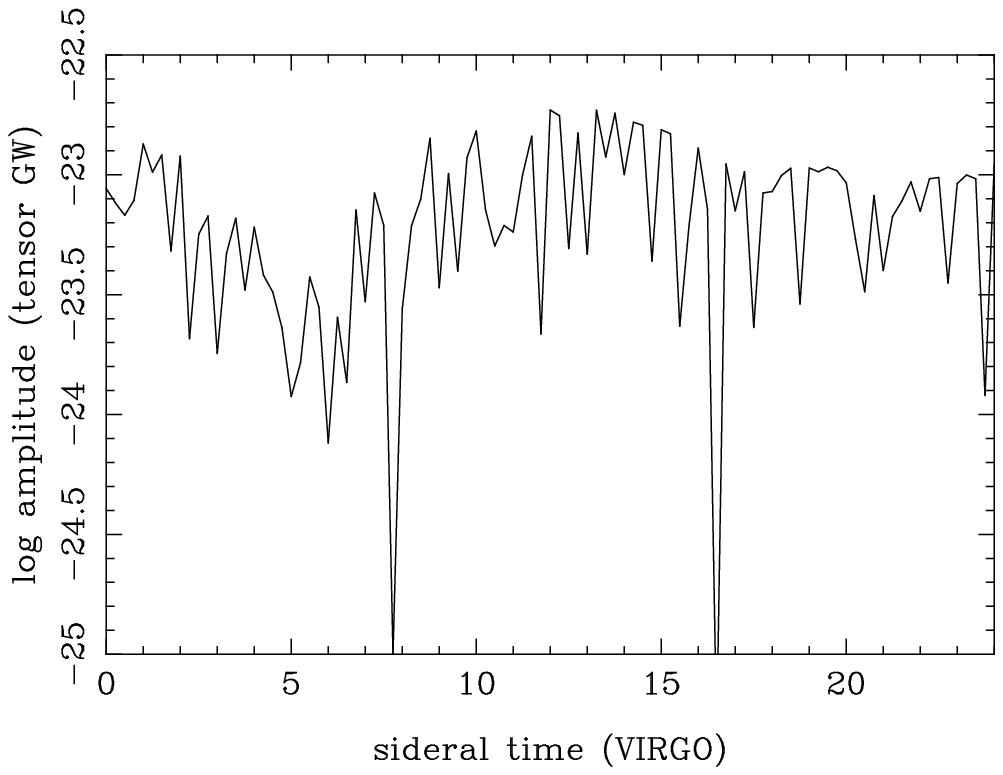}}
\resizebox{\hsize}{!}{\includegraphics{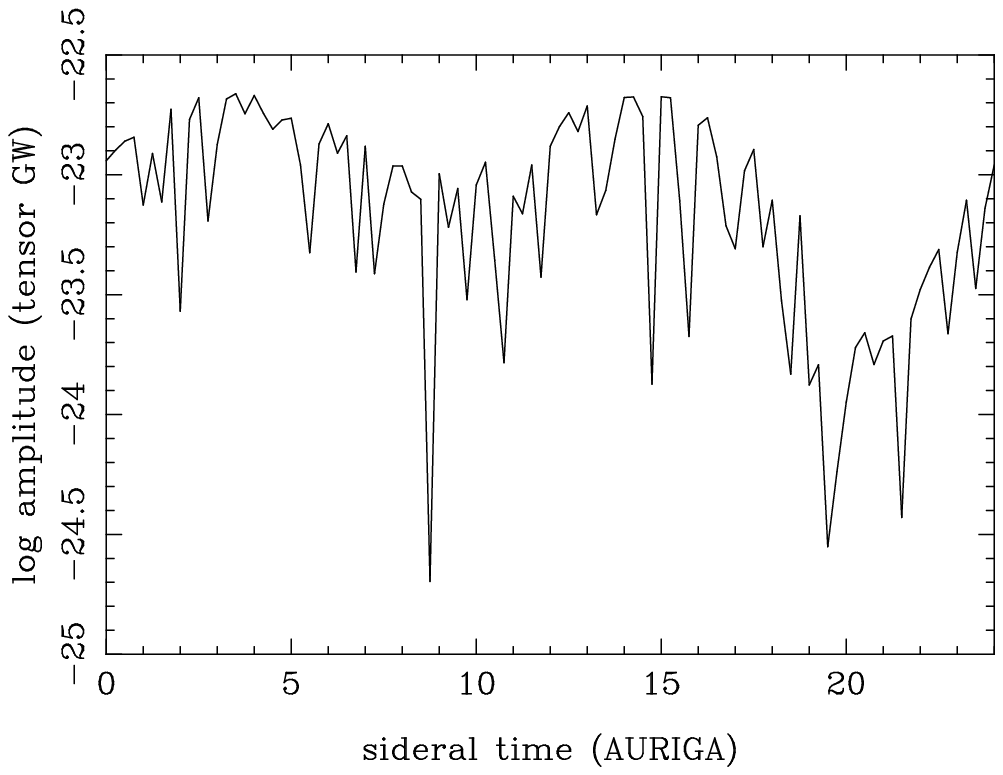}}
\caption{a) Amplitude for the VIRGO detector as in Fig.\ref{ampliTIi}-a but with
random directions of polarization.
The shape is unchanged but some events have a reduced amplitude.
This will lead to a reduction of the GW events detected at a given sensitivity level.
b) Amplitude for the AURIGA detector as in Fig.\ref{ampliTIi}b but with
random directions of polarization.
The shape is unchanged but, on average, the amplitude is reduced.}
\label{ampliTIa}
\end{figure}

For interferometric detectors, 
the different curves are shifted along the x-axis (sidereal time) depending on the
polarization. On the contrary, for bar detectors, the curves are shifted along the
y-axis (amplitude), but the minima and maxima always appear at the same x
values (same sidereal times).
Let us explain why it is not correct to calculate the mean over the
different polarizations. If an event is produced with the favorable polarization,
it will be detected if the amplitude is larger than the limiting amplitude. On the other hand,
if the polarization is unfavorable, the observed amplitude will be reduced and
the considered event may fall below the limiting amplitude. The global effect of
the uncertainty on the polarization is simply a reduction of the number of counted
events, but not a reduction of the amplitude.
Finally, the distribution of the amplitudes along the sidereal time will be simply given by
the envelope of the curves obtained for the different polarizations.
It must be noted that it is a handicap for interferometric detectors because the
contrast is smoothed and the total number of expected events is reduced.
We repeated the same calculations with polarizations taken at random
(Fig. \ref{ampliTIa}a-b). The same effect is clearly visible. The curves
are the envelopes of the previous ones. Some events are seen with smaller 
amplitudes due to unfavorable polarization.

\subsection{Comparison of the amplitudes of GW from Virgo and the Great Attractor}

We have calculated the GW amplitudes as a function of sidereal time
for the four detectors listed in the previous table.
For each of them we give two figures, for tensor and scalar GW, respectively.
In each figure we present the amplitudes expected for sources in the
Virgo cluster (solid curve) and in the Great Attractor region (dashed curve).

\begin{figure}[!]
\resizebox{\hsize}{!}{\includegraphics{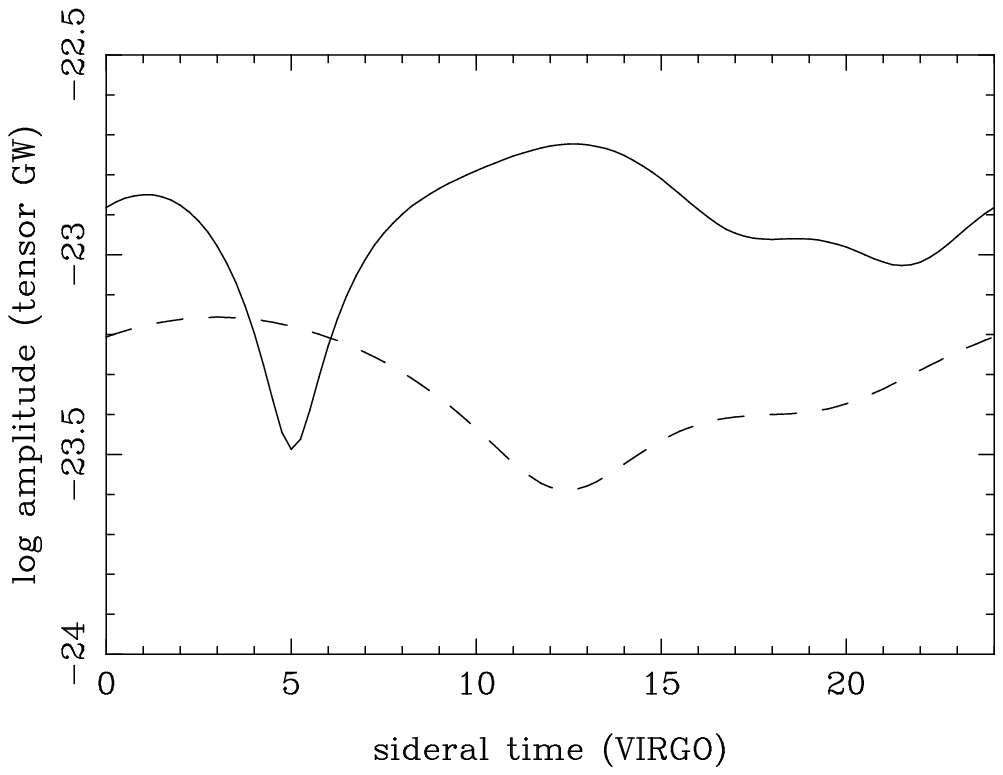}}
\resizebox{\hsize}{!}{\includegraphics{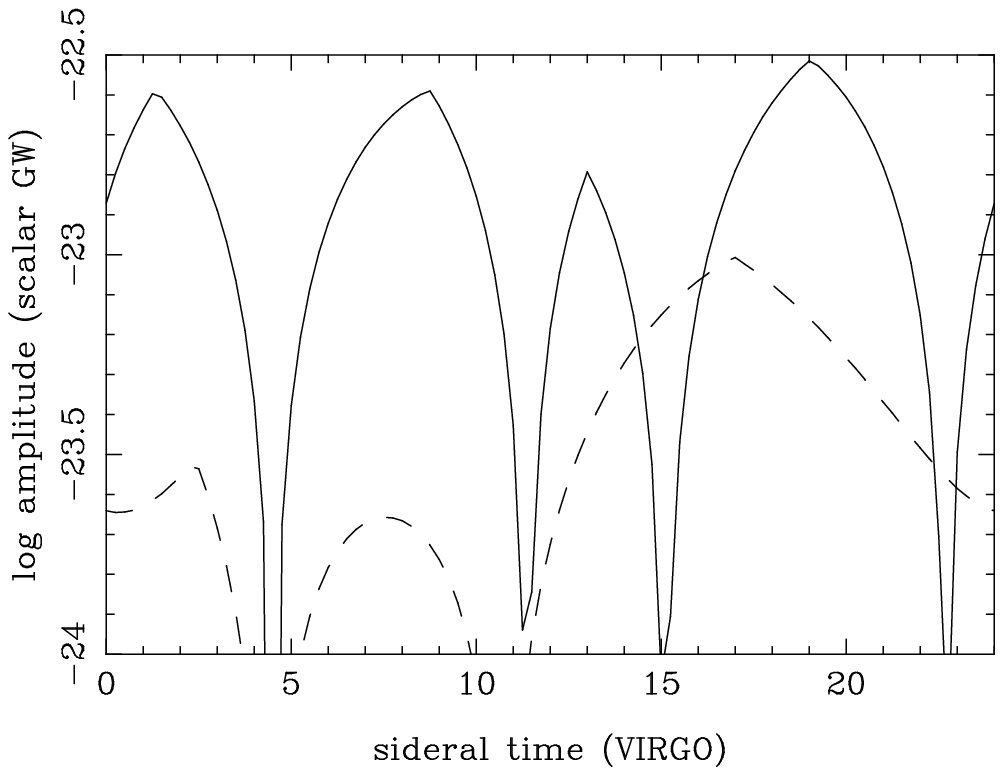}}
\caption{a) Amplitude as a function of sidereal time for tensor GW emitted
by Virgo (solid curve) and the Great Attractor (dashed curve) as seen with
the interferometric VIRGO detector. With a sensitivity of $10^{-23  }$ only
Virgo will give large enough amplitude to make events detectable between
$t=9h$ and $t=15h$ and between $23h$ and $4h$.
The Great Attractor could be detected with a
sensitivity of $10^{-23.5}$ (except between $t=11h$ and $t=15h$).
b) Ibidem but for scalar waves. Virgo (solid curve)
could be detectable at different peaks along the sidereal time.
The Great Attractor could be barely detected with a sensitivity of $10^{-23  }$ at $t=17h$ }
\label{ampliTIm}
\end{figure}

\begin{figure}[!]
\resizebox{\hsize}{!}{\includegraphics{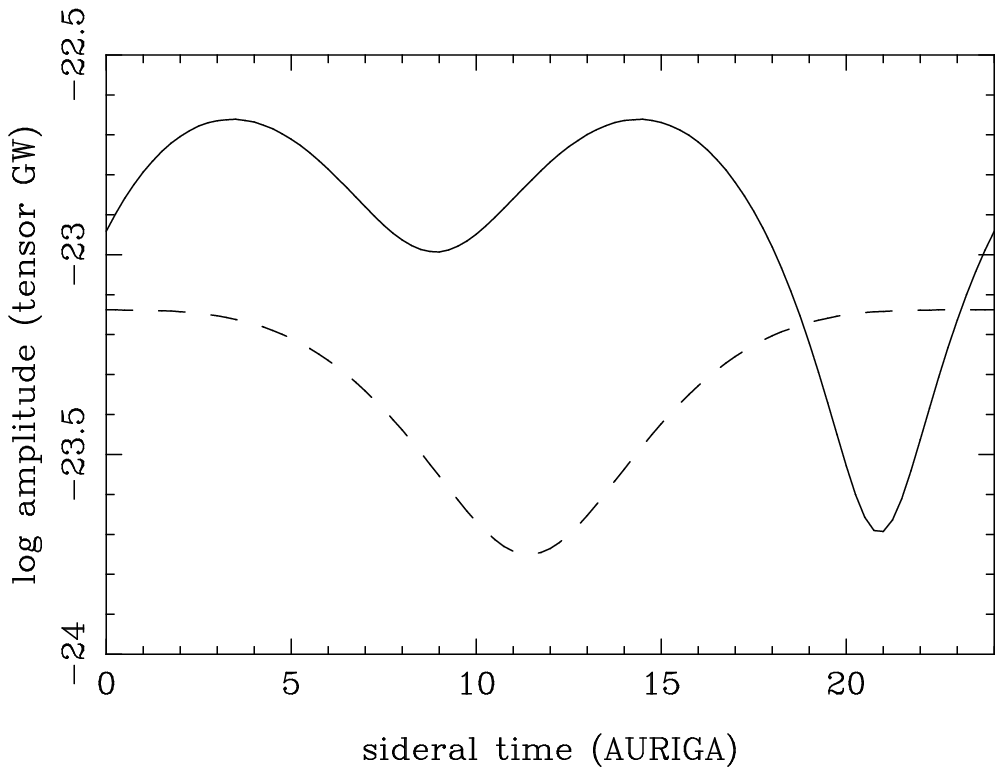}}
\resizebox{\hsize}{!}{\includegraphics{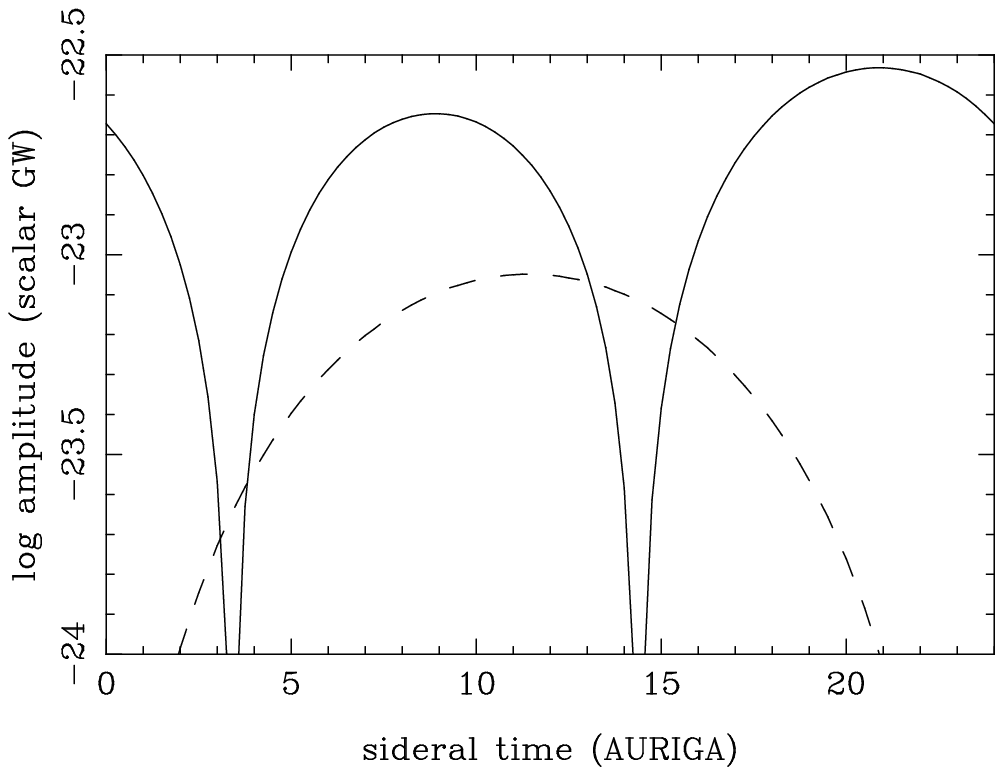}}
\caption{a) Amplitude as a function of sidereal time for tensor GW emitted
by Virgo (solid curve) and the Great Attractor (dotted line) as seen with
the bar detector AURIGA. With a sensitivity of $10^{-23  }$
Virgo will be detectable between
$t=0h$ and $t=18h$.
b) Ibidem for scalar waves. Virgo could
be detectable at two main positions $t=9h$ and $t=21h$.
The Great Attractor could be barely detected with a sensitivity of $10^{-23  }$
at $t=12h$. }
\label{ampliTB1m}
\end{figure}

\begin{figure}[!]
\resizebox{\hsize}{!}{\includegraphics{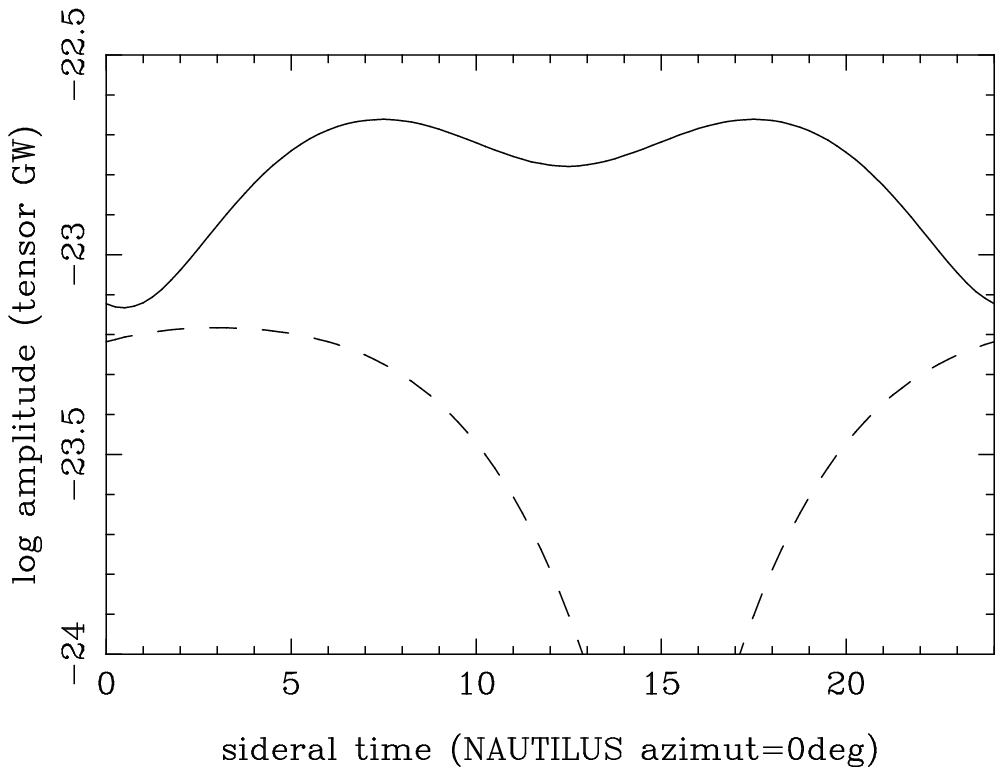}}
\resizebox{\hsize}{!}{\includegraphics{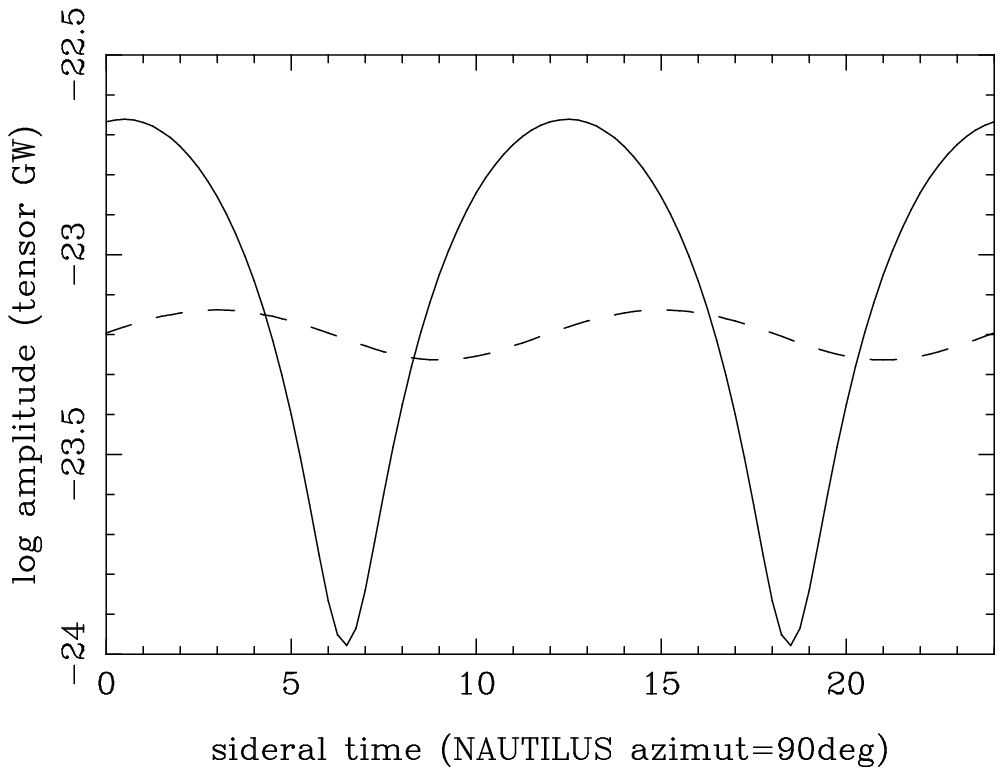}}
\caption{a) Amplitude as a function of sidereal time for tensor GW emitted
by Virgo (solid curve) and the Great Attractor (dashed curve) as seen with
the bar detector NAUTILUS oriented in the direction of the north (azimuth=$0\deg$).
This figure is the same as Fig.\ref{ampliTB1m}-a but with a shift in sidereal time.
b) Ibidem with NAUTILUS oriented in the east-west
direction (azimuth=$90\deg$). Virgo could be detectable at two main positions $t=1h$
and $t=12h$. Figures a and b have opposite phases.  }
\label{ampliTB2m}
\end{figure}

\begin{figure}[!]
\resizebox{\hsize}{!}{\includegraphics{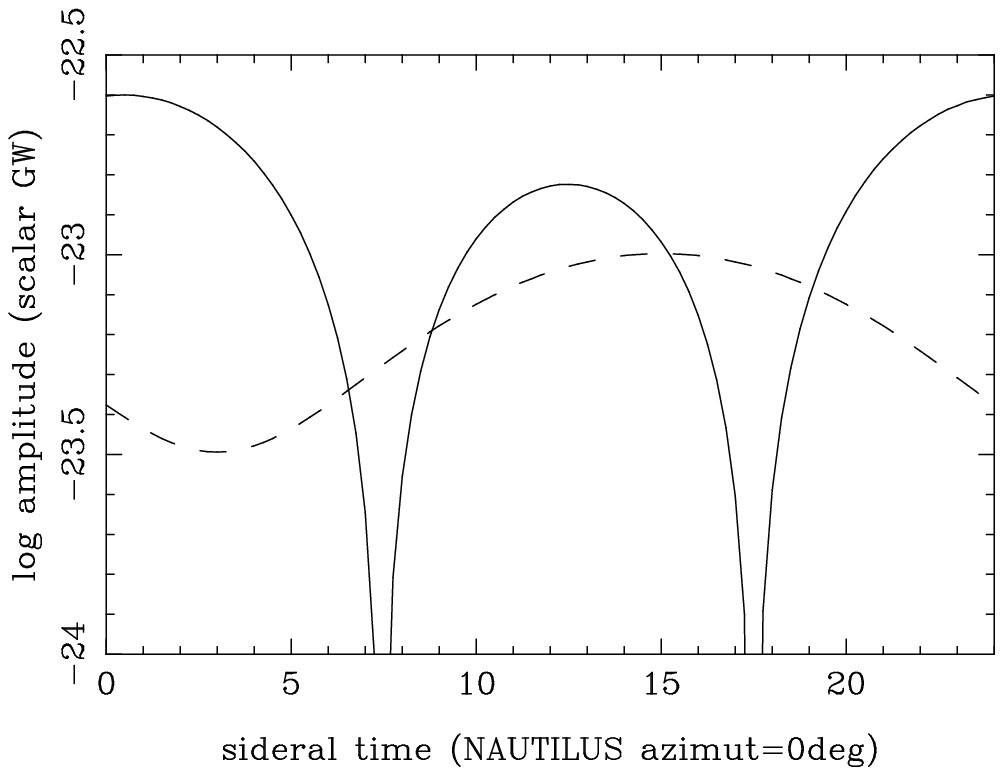}}
\resizebox{\hsize}{!}{\includegraphics{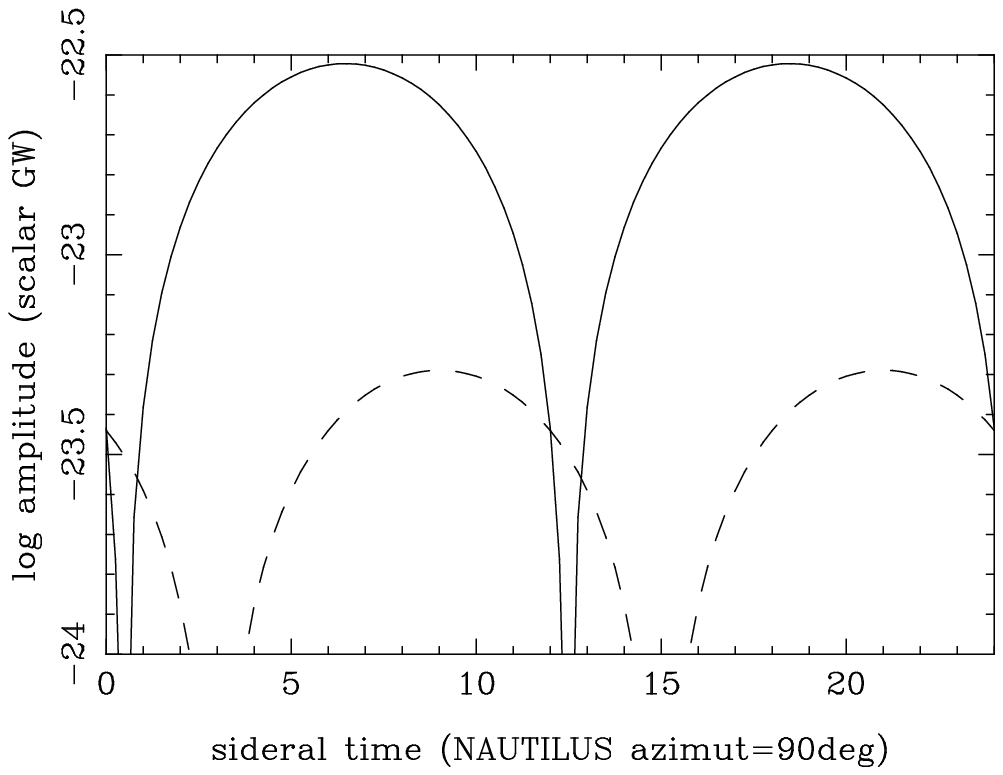}}
\caption{a) Amplitude as a function of sidereal time for scalar GW emitted
by Virgo (solid curve) and the Great Attractor (dashed curve) as seen with
the bar detector NAUTILUS oriented in the direction of the north (azimuth=$0\deg$).
This figure is the same as Fig.\ref{ampliTB2m}-a but with a shift in sidereal time.
b) Ibidem with NAUTILUS oriented in the east-west
direction (azimuth=$90\deg$). Virgo could be detectable at two main positions $t=7h$
and $t=18h$. Again, Figures a and b have opposite phases.}
\label{ampliSB2m}
\end{figure}

The results are given in four figures (Fig. \ref{ampliTIm} to \ref{ampliSB2m}):
The caption of each figure contains detailed comments.
Here, we will simply
highlight prominent features.

With a sensitivity $h_{lim}=10^{-23  }$ only GW events from the Virgo
cluster (solid lines) will be detectable. The Great Attractor (dashed curves)
will be detectable only with a sensitivity of $h_{lim}=10^{-23.5}$.
Nevertheless, there is one case (VIRGO detector - Fig.\ref{ampliTIm}b)
where scalar waves
could be seen from the Great Attractor with a sensitivity of $h_{lim}=10^{-23  }$.
Because we expect about 25 times more events from the Great Attractor,
this may result in a peak in the rate of events around sidereal time
$t=17h$.

Another important features in these diagrams of amplitudes is
that tensor and scalar waves give peaks which generally have opposite phases.
In other words, if we expect a maximum of events for scalar waves, there should be
a minimum for tensor waves. This is clearly visible by comparing
Fig.\ref{ampliTB1m} a and b. This is also important because
it may be used to disentangle the contributions of these two kinds of waves.
The same characteristic is present when we compare the expected counts for
the NAUTILUS bar detector with two perpendicular orientations ($\Phi_o=0 \ \deg$ and
$\Phi_o=90 \ \deg$). The figure suggests that NAUTILUS could benefit from an orientation
complementary to the one used, e.g., with the AURIGA orientation.

\subsection{Number of GW events expected for the actual galaxy distribution within 100 Mpc}

\begin{figure}[p]
\resizebox{\hsize}{!}{\includegraphics{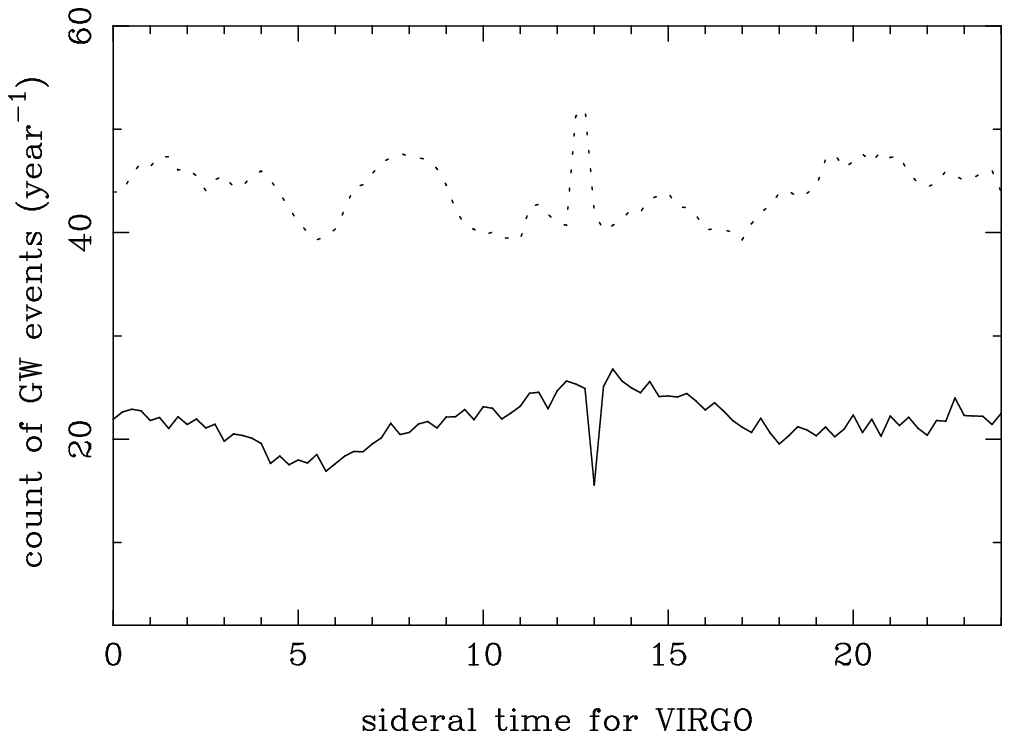}}
\resizebox{\hsize}{!}{\includegraphics{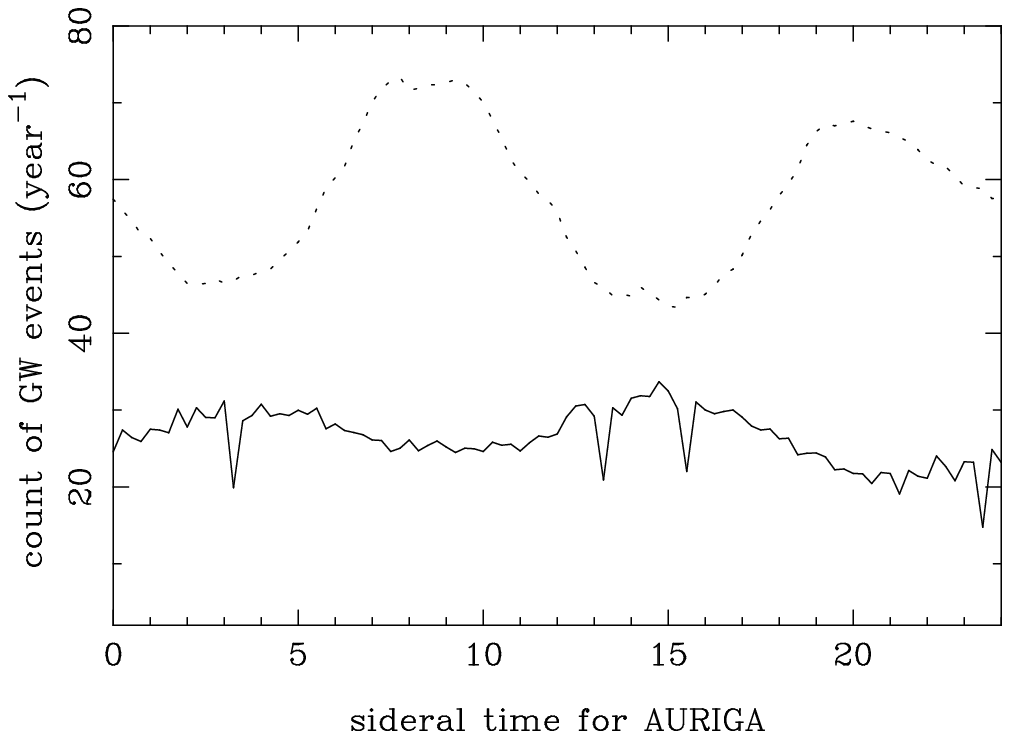}}
\caption{
a) Density of probability of GW events (number per year) for the VIRGO
interferometric detector as a function of the sidereal time of the site.
The calculation is made from the actual sample of galaxies described in
section 2.
Tensor waves are shown with a solid curve, while scalar waves are
given as dashed curves. These predicted counts are obtained from Eq.\ref{proba}
using case b of Fig.\ref{sens_dist} for the calculation of the observed
amplitude. The sensitivity is assumed to be $h_{lim}=10^{-23  }$.
Because one retrieves the expected distribution found for Virgo alone
(solid curves of Fig.\ref{ampliTIm}), one
concludes that the Virgo cluster will be the dominant source of GW events
with such a sensitivity.
b) Ibidem for the AURIGA bar detector. The comparison
with predicted amplitude for Virgo cluster alone also confirms that it will be
the dominant source of GW events at the considered sensitivity. It is clearly
visible that tensor and scalar waves have opposite phases.  }
\label{COUNT_A}
\end{figure}

\begin{figure}[p]
\resizebox{\hsize}{!}{\includegraphics{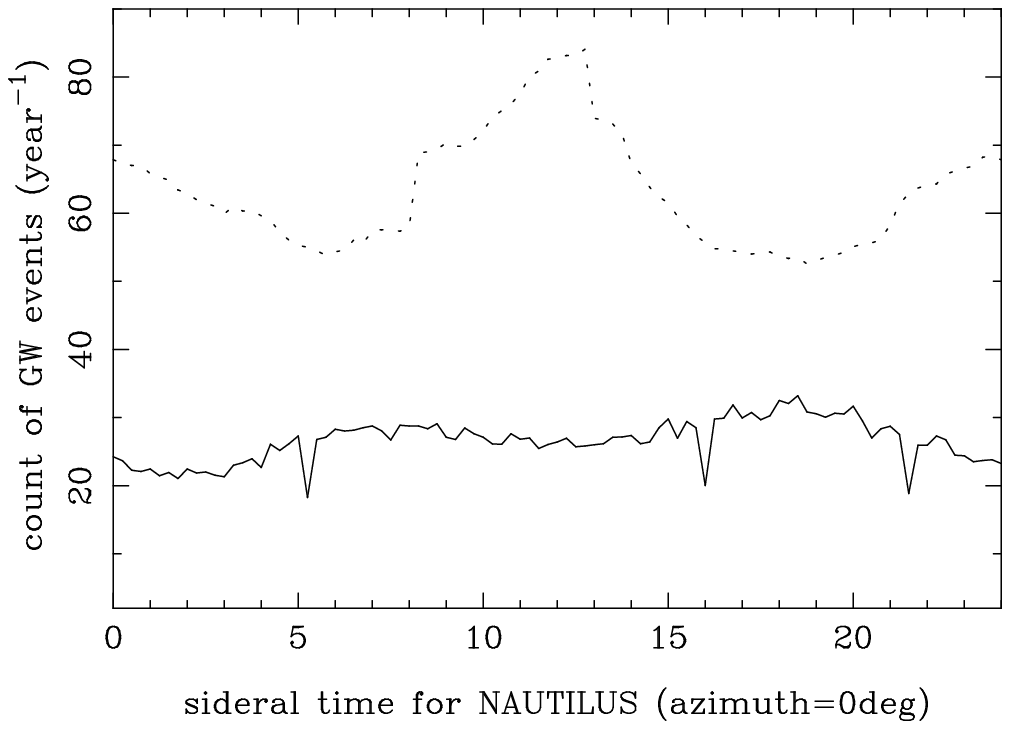}}
\resizebox{\hsize}{!}{\includegraphics{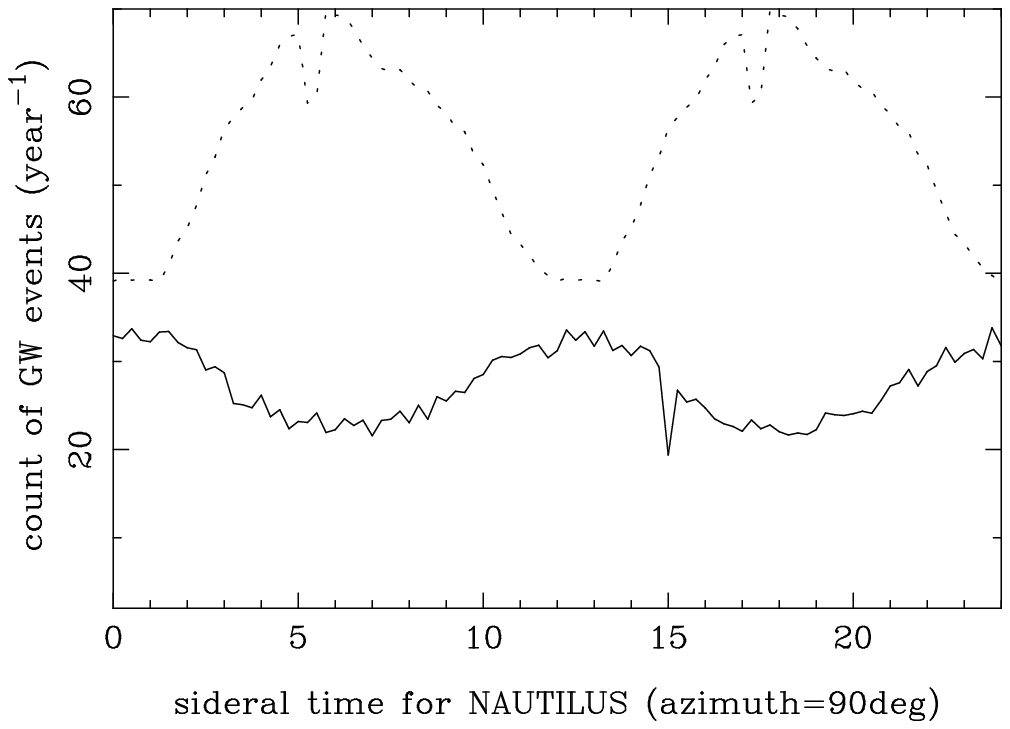}}
\caption{a) Same as the previous figure for the NAUTILUS pointing towards the
north ($\Phi _o=0 \ \deg$). The results and the conclusions are very similar
to those obtained from Fig.\ref{COUNT_A}-~b. Nevertheless we note that
this orientation is less favorable for scalar waves in comparison with
the AURIGA bar detector.
b) Ibidem for the NAUTILUS pointing towards the
west ($\Phi _o=90 \ \deg$).  We note the improvement of detected GW events
when comparing with the previous orientation.
}
\label{COUNT_C}
\end{figure}

Using the catalog of galaxies described in section 2,
we simulated the count of GW events using Eq. \ref{proba}.
The calculation is made with an energy of $E_{gw}= 10^{-6} M_{\odot}c^2$,
a frequency of $1 \ kHz$ and a duration of $1 \ ms$. 
This corresponds to the case b
of Fig.\ref{sens_dist}. The adopted limiting amplitude $h_{lim}$ is
$10^{-23  }$.  It should allow us to reach the Virgo cluster.
The calculation is done for VIRGO, AURIGA, NAUTILUS (azimuth=$0 \deg$)
and NAUTILUS (azimuth=$90 \deg$).
We plotted simultaneously the counts for tensor (solid curves) and scalar waves
(dashed curves).
The results are shown in Fig.\ref{COUNT_A} and \ref{COUNT_C}
for the four considered detectors.
When comparing these Figures with the expected amplitudes calculated
for the detectors VIRGO and AURIGA, it is seen
that all predicted maxima are at the expected positions 
according to  the distribution
of amplitudes given by the Virgo cluster alone 
(solid curves in Fig. \ref{ampliTIm}a and b and \ref{ampliTB1m}a and b).
This confirms that, within a
distance of 20 Mpc, the Virgo cluster dominates. This was not obvious because
the influence of other galaxies was not easy to predict.
The expected number of GW events reaches a few tens per year at the most
favorable sidereal time but with a yet unreachable sensitivity
for the considered GW energy.
For instance, from Fig.\ref{COUNT_A}-a we see that several maxima are
expected for scalar waves (dotted curve) in agreement with their positions predicted
from Fig.\ref{ampliTIm}-b for the Virgo cluster alone. Similarly, the maximum
at $t \approx 13h$ for tensor waves (solid curve) is the one predicted for Virgo alone as
seen from Fig.\ref{ampliTIm}-a.
This confirms that the Virgo cluster will be the dominant source of GW events
when the sensitivity will be $h_{lim}=10^{-23  }$.

From Fig.\ref{COUNT_A}-b (AURIGA bar detector)
one can see clearly that tensor waves
and scalar waves have opposite phases as far as the sidereal time is concerned.
The same effect is also visible from Fig.\ref{COUNT_C} a and b
with the NAUTILUS bar detector. Further, we see that changing the
orientation of the bar by $90 \ \deg$ also produces a change of sidereal time
phase; The maximum in Fig.\ref{COUNT_C}-b for, say, tensor waves (solid curve),
corresponds to a minimum in Fig.\ref{COUNT_C}-a.

\section{Discussion and conclusions}

From the detection of GW events with bar and interferometric detectors, 
one can see that, due to the different geometrical factors 
and to the anisotropy of the distribution of the GW sources,
it is in principle possible to make the distinction between 
transversal tensor GW,
transversal scalar GW and longitudinal scalar GW.

To demonstrate this difference we calculated the expected amplitudes and
the number of events as a function of sidereal time
for real positions of existing GW detectors
and for real 3-dimensional distribution of galaxies within 100 Mpc.
If one adopts the  value for the 
energy of GW pulse 
$E_{gw} = 10^{-6} M_{\odot} c^2$ (lines $b$ in Fig.\ref{sens_dist}), 
then, GW events produced
at the distance of the Virgo cluster can be detected only
with a sensitivity, yet unreachable, of $h_{lim}=10^{-22.5}$.
However if the GW energy is about 
$E_{gw} = 10^{-3}M_{\odot}c^2$ (case $a$ in Fig.3)
as predicted by nonaxisymmetric scenarios of SN core collapse,
then Virgo cluster and
Great Attractor would be visible with $h_{lim} = 10^{-22}$.

 If the GW energy emission has value
 $E_{gw} = 10^{-9} M_{\odot} c^2$ or less,
as predicted by an axisymmetric scenario of SN explosion,
then Virgo would simply be not detectable with present
and forthcoming detectors. Thus, one cannot expect a high
detection rate, but only serendipitous detections from
very nearby SN's.

We would like to emphasize a point which seems important
to us. Today there are several scenarios of GW radiation from
SN core collapse even in the frame of the General Relativity,
which predicts only tensor gravitational radiation
(see review by Thorne,1997).
Within different scenarios, the SN core collapse may
lead to a large range of radiated GW energy
from $10^{-11} M_{\odot}c^2$ up to $10^{-2} M_{\odot}c^2$
and to very
different forms of GW bursts, i.e. to the different
spectral energy distributions and durations of GW pulses
from msec up to sec and even minutes (Lai \& Shapiro, 1995).
It is important to recall
that other relativistic and quantum gravity theories
(such as string theory, the Jordan-Fierz-Brans-Dicke theory and
tensor field theory) predict scalar gravitational
radiation which is generated also in the case of a spherical
gravitational collapse and  which may
carry a large GW energy.

We would like to emphasize the importance of considering a
wide range of GW burst parameters for SN core collapses
by discussing the SN1987A and SN1993J events.
Analysis of data
recorded by Geograv for SN1987A (Amaldi et al. 1987) and
by Explorer-Allegro for SN1993J (Mauceli et al. 1997),
showed that there are GW candidate events,
which the authors themselves do not consider as real signals
because the GW energy calculated for a standard pulse
with a duration $\tau_g = 1 ms$ (and hence bandwidth 1 kHz)
gives $E_{gw} \approx 10^3 M_{\odot}c^2$ for both supernovae.
However,  in the case of pulse duration of about 1 sec 
(and hence bandwidth 1 Hz)
the GW energy
needed for producing the same GW amplitude is about
$1 \ M_{\odot}c^2$. In this case the observed GW amplitudes
correspond to the lines ``a'' in Fig.3 and fit well the
decrease expected from the
relative distances of the two host galaxies (LMC, M81)
\footnote{The high value of energy release in a
gravitational collapse could also be supported by the existence of
gamma-ray bursts if they are related to SN explosions because
they may generate in the electromagnetic channel an energy
of about 1 $M_{\odot}c^2$ with timescale a few seconds
 (Kulkarni et al. 1999; Paczynski 1999).}.

This means that GW data analysis should be done for the interval
of possible signal durations from  ms to  sec timescales.
For the GW pulses longer than 1 sec
the frequency bandwidth is less than 1 Hz and the sensitivity
of bar detectors may be essencially improved if one uses as
a signal the difference between two signals coming from
two resonances of a bar detector.

\medskip
Let us conclude with the most secure results presented in this
paper:

\begin{itemize}
\item We wrote a computer code to calculate the amplitude of
transversal or longitudinal gravitational waves from any real source
which could be detected with bar or interferometer detectors.
This code allows us to fix all experimental conditions: latitude
and sideral time of the detector, type and orientation of the detector,
right ascension and declination of the emitting source, characteristics
of the gravitational wave.

\item This code has been used to calculate how the amplitude
changes with sideral time of the site for existing detectors.
In particular, using a point source we study
the effect of the polarization of the wave in the case of tensor
waves.

\item Then, we compared the distribution of amplitudes along the
sideral time for bar and interferometric detectors and for transversal
and longitudinal gravitational waves for two dominant point sources.
We found an interesting result for bar detectors (NAUTILUS) which can 
have different orientations.  This may be used for the selection of 
real GW signals from the noise.
The result is that it would be theoretically  possible to discriminate 
both kinds of waves because, for a considered point source, the maxima 
of the distributions do not appear at the same sideral time.

\item Finally, we applied the code to calculate the expected count
of events generated by the actual distribution of galaxies
using the gravitational energy release predicted by existing scenarios 
of Supernova core collapse. The result is that, with the sensitivity
of GW detectors $h = 10^{-22}$ and for the released GW energy 
$E_{gw} = 10^{-3}$, predicted by nonaxisymmetric scenarios
of SN core collapse supernova, it is possible to detect
GW events from the distance of the Virgo cluster and Great Attractor
and to use the statistics of the events as a test of gravity theories.

\end{itemize}

\appendix
\section{The longitudinal character of the scalar wave
in the field gravity theory}

Feynman's field gravity theory (Feynman et al.,1995)
is based on the Lagrangian formalism of
the relativistic quantum field theory
and presents a non-geometrical description
of gravitational interaction.
According to Feynman, the gravity force between two masses
is caused by the exchange of gravitons
which are mediators of the
gravitational interaction and actually represent the quantum
of the relativistic tensor field $\psi^{ik}$ in Minkowski space
$\eta^{ik}$.

It is important that the problem of the physical interaction
of a gravitational wave with an detector
may be completely analyzed in terms of the weak field approximation
where classical relativistic field theory is applicable
(see Landau \& Lifshitz, 1971).

First, let us consider the general problem of
the motion of a relativistic test particle having
rest-mass $m_0$, 4-velocity $u^i$, and 3-velocity $\vec{v}$
in the gravitational field described by the symmetric
tensor potential $\psi^{ik}$ in the flat Minkowski space-time.
The Cartesian coordinates always exist and the metric tensor is
$\eta^{ik}=diag(1,-1,-1,-1)$
(we utilize notations
of the text-book Landau \& Lifshitz, 1971 ).

To derive the equation of motion for test particles
in the frame of the field gravity theory
we start from the stationary action principle in the form
\begin{equation}
\label{delta-S}
\delta S = \delta ( \frac{1}{c} \int (\Lambda_{(p)} + \Lambda_{(int)})
d \Omega ) = 0
\end{equation}
where the variation of the action is made with respect to the
particle trajectories $\delta x^i$
for fixed gravitational potential $\psi^{ik}(t,\vec{r})$,
and $d\Omega$ is the element of the 4-volume.

The free particle Lagrangian is
\begin{equation}
\label{Lambda-p}
\Lambda_{(p)} = -\eta_{ik} T_{(p)}^{ik}
\end{equation}
and the interaction Lagrangian is
\begin{equation}
\label{Lambda-i}
\Lambda_{(int)} = -\frac{1}{c^2}\psi_{ik} T_{(p)}^{ik}
\end{equation}
where the energy-momentum tensor of the test point particle is
\begin{equation}
\label{Tpik}
T_{(p)}^{ik}=m_0 c^2\delta(\vec{r}-\vec{r_p})\{1-\frac{v^2}{c^2}\}^{1/2}
u^i u^k
\end{equation}
The result of variation gives the equations of motion in the form
(Kalman 1961; Baryshev 1986):
\begin{equation}
A_k^i \frac{dp^k}{ds}=-m_0 B^i_{kl}u^k u^l
\label{eq-motion}
\end{equation}
where $p^k = m_0c u^k$ is 4-momentum of the test particle,
$(\cdot)_{,i} = \frac{\partial (\cdot)}{\partial x^i} $   and
\begin{equation}
A_k^i=(1- \frac{1}{c^2}\psi_{nl}u^n u^l)\eta_k^i -
\frac{2}{c^2}\psi_{kn}u^n u^i + \frac{2}{c^2}\psi_k^i
\label{Aik}
\end{equation}
\begin{equation}
B_{kl}^i= \frac{2}{c^2}\psi_{k,l}^i
-\frac{1}{c^2}\psi_{kl}^{\;\; ,i}
-\frac{1}{c^2}\psi_{kl,n}u^n u^i
\label{Bikl}
\end{equation}
In Eq.(\ref{eq-motion}) the rest mass of the test particle
may be canceled, hence in the
field gravity theory $m_{in}=m_g=m_0$ without the initial equivalence
postulate (Thirring 1961). In agreement with Eq.(\ref{eq-motion})
all classical relativistic post-Newtonian gravity effects
have the same values as in general relativity (see e.g. Baryshev 1995).

The scalar gravitational wave in the field gravity theory
is generated by the trace of the energy momentum
tensor $T=\eta_{ik}T^{ik}$
of the sources of the gravitational potential
(Baryshev 1982;1995; Sokolov 1992)
and may be expressed in the form:
\begin{equation}
\psi^{ik}(t,\vec{x})=\frac{1}{4} \psi(t,\vec{x})\; \eta^{ik}
\label{scalar}
\end{equation}
where $\psi=\eta_{ik}\psi^{ik}$ is the trace of the tensor
potential $\psi^{ik}$ and hence is a 4-scalar field
in Minkowski space.

Substituting Eq.(\ref{scalar}) in Eq.(\ref{eq-motion})
and taking into account the weak field approximation
we get the following equation of motion of the test particle
in scalar gravitational potential:
\begin{equation}
\frac{du^i}{ds}=
\frac{1}{4c^2}\psi^{,i} -\frac{1}{4c^2} \psi_{,l}u^l u^i
\label{weak_scalar_eq_motion}
\end{equation}

Spatial components of this equation ($i=\alpha$) give the expression
for the gravity force acting on the test particle:
\begin{equation}
\frac{d\vec{p}}{dt}=
-\frac{m_0}{4 \sqrt{1-\frac{v^2}{c^2}}}
\lbrack ( 1-\frac{v^2}{c^2}) \vec{\nabla}\psi +
\frac{\vec{v}}{c} ( \frac{\partial \psi}{c \partial t} +
\frac{\vec{v}}{c}\cdot \vec{\nabla} \psi ) \rbrack
\label{scalar_force}
\end{equation}
The corresponding 3-acceleration of the test particle is
\begin{equation}
\frac{d\vec{v}}{dt}=
-\frac{1-\frac{v^2}{c^2}}{4}
(\vec{\nabla}\psi +
\frac{\vec{v}}{c}\frac{\partial \psi}{c\partial t})
\label{scalar_accel}
\end{equation}
The time component ($i=0$) of the Eq.(\ref{weak_scalar_eq_motion})
gives the work produced by the scalar gravitational wave:
\begin{equation}
\frac{dE_k}{dt}=
-\frac{m_0}{4 \sqrt{  1-\frac{v^2}{c^2}  }}
(\vec{v}\cdot\vec{\nabla}\psi +
\frac{v^2}{c^2}\frac{\partial \psi}{c\partial t})
\label{scalar_work}
\end{equation}
where $E_k$ is the kinetic energy of the particle.
For the case of slow motion of the test particle ($v/c \ll 1$)
we get for the gravity force:
\begin{equation}
\frac{d\vec{p}}{dt}=
-\frac{m_0}{4}
\vec{\nabla} \psi
\label{slow_scalar_force}
\end{equation}
for 3-acceleration of the test particle:
\begin{equation}
\frac{d\vec{v}}{dt}=
-\frac{1}{4}
\vec{\nabla}\psi
\label{weak_scalar_accel}
\end{equation}
and for the work of the gravity force:
\begin{equation}
\frac{dE_k}{dt}=
-\frac{m_0}{4}
(\vec{v}\cdot\vec{\nabla}\psi)
\label{weak_scalar_work}
\end{equation}

The scalar plane monochromatic GW  in the system of coordinates
with the z-axis directed along the wave propagation may be written in
the form:
\begin{equation}
\psi_{ik} =\frac{1}{4} \psi(t,z)\;\eta_{ik}
\end{equation}
\begin{equation}
\psi_{ik} 
= h_0 \cos(\omega t - kz)\; diag(1,-1,-1,-1)
\label{scal}
\end{equation}

where $h_0\ll c^2$ is the amplitude of the wave,
$\omega = 2\pi \nu$, $k =\omega/c = 2\pi /\lambda$.

For the scalar gravitational potential Eq.(\ref{scal})
the 3-acceleration Eq.(\ref{weak_scalar_accel}) of the test particle is
\begin{equation}
\frac{dv_x}{dt}=\frac{dv_y}{dt}=0
\label{vxy}
\end{equation}
\begin{equation}
\frac{dv_z}{dt}=\frac{1}{4c}\frac{\partial \psi}{\partial t}
\label{vz}
\end{equation}
According to Eq.(\ref{vz})  under the action
of the scalar gravitational wave a test mass
undergo small longitudinal oscillations in the direction
of wave propagation (z-axis):
\begin{equation}
\label{z(t)}
z(t)= \frac{\lambda}{2\pi}\frac{h_0}{c^2} sin(\omega t - kz_0) +z_0
\end{equation}
The relative distance between two particles
$\Delta z(t) = z_1(t) - z_2(t) = l(t)$
gives the relative oscillations $\Delta l(t) = l(t) - l_0 =
\Delta l_0 \; cos(\omega t +\alpha)$, which may be detected
by gravitational detector.

The very important property of the scalar gravitational potential
is that it does not interact with the electromagnetic
field. Indeed, the interaction Lagrangian for the potential
Eq.(\ref{scalar}) is
\begin{equation}
\label{Lambda_iem}
\Lambda_{(int)} = -\frac{1}{c^2}\psi_{ik} T_{(em)}^{ik} =0
\end{equation}
because the trace of the energy momentum tensor of the
electromagnetic field $T_{(em)} = 0$. This means that the above
analysis is sufficient for the study of the response of the
bar or interferometric detectors where the distances between
test particles are measured by means of the electromagnetic field.

For a  detector with the length $l_0 \ll \lambda$
the amplitude of the relative displacement $\Delta l_0 / l_0$
between two test particles, which is  measured by the detector is
\begin{equation}
\label{dl/l}
\frac{\Delta l_0}{l_0} =  h_0 \;cos\theta
\end{equation}
where $h_0$ is the amplitude of the scalar wave Eq.(\ref{scal}),
and $\theta$ is  the angle between the direction of the wave
propagation and the line connecting the two masses.

For a bar detector the integral cross section
for the scalar gravitational wave  having $\Delta \nu_g >
\Delta \nu_a $
is given by (Baryshev, 1997):
\begin{equation}
\label{sigma}
\sigma_{(scalar)}=
\frac{8}{\pi}\frac{G}{c^3} M V_s^2 \;cos^2 \theta
\end{equation}
where $M$ is the mass of the cylinder, $V_s =\omega_0 L / \pi $
is the speed of sound in the cylinder, $\omega_0$ is resonance
angular frequency of the bar, $L$ is the length of the cylinder
and $\theta$ is the angle between the bar axis and the direction
of the wave propagation. The factor $cos\theta$
in Eqs.(\ref{dl/l},\ref{sigma})  shows the
longitudinal character of the scalar gravitational wave.

\acknowledgements{
We thank
Giovanni Pallottino and Paolo Bonifazi for giving parameters of bar
detectors and comments, Pekka Teerikorpi and Vladimir Sokolov
for useful discussions which helped to improve the text.
Y.B. thanks the President of the University of Lyon I
for inviting him on a temporary position.
We thank the anonymous referee for usefull coments.
}


\end{document}